\documentclass[useAMS,usenatbib]{mn2e}

\usepackage[usenames,amssymb,dvipsnames]{xcolor}
\usepackage{epsfig}
\usepackage{graphics}
\usepackage{psfig}
\usepackage{placeins}
\usepackage{flushend}
\usepackage{cuted}
\usepackage{widetext}
\usepackage{bm}
\usepackage{times}
\usepackage{amssymb}
\usepackage{caption}
\usepackage{tipa}
\usepackage{amsmath}
\usepackage{times}

\newcommand\be{\begin{equation}}
\newcommand\ee{\end{equation}}
\newcommand\bea{\begin{eqnarray}}
\newcommand\eea{\end{eqnarray}}
\newcommand{\cfl}      {\ensuremath{~\mathcal{C}_{\rm CFL}}}

\def\gsim{ \lower .75ex \hbox{$\sim$} \llap{\raise .27ex \hbox{$>$}} }
\def\lsim{ \lower .75ex\hbox{$\sim$} \llap{\raise .27ex \hbox{$<$}} }

\title[GR-RHD of accretion flows: II. Employing a stiff solver]
{General relativistic radiation hydrodynamics of accretion
  flows: II. Treating stiff source terms and exploring physical limitations%
}

\author[C.~Roedig, O.~Zanotti, D.~Alic]{C.~Roedig\thanks{croedig@aei.mpg.de}$^{1}$, O.~Zanotti$^{2}$, D.~Alic$^{1}$ \\
$^{1}$ Max-Planck-Institut f{\"u}r Gravitationsphysik, Albert Einstein 
Institut, Am M\"uhlenberg 1, 14476 Golm, Germany \\
$^2$ Universit\`a di Trento, Laboratorio di Matematica Applicata, Via Messiano 77, I-38100 Trento, Italy
}

\begin{document}

\date{\today }

\maketitle

\label{firstpage}

\begin{abstract}
We present the implementation of an
\textit{\textbf{im}plicit-\textbf{ex}plicit} (IMEX) Runge-Kutta
numerical scheme for general relativistic hydrodynamics coupled to
an optically thick radiation field in two existing GR-(magneto)hydrodynamics codes.
 We argue that the necessity of such an improvement arises naturally
in most astrophysically relevant regimes where the
optical thickness is high as the equations become stiff. 
By performing several simple one dimensional tests we
verify the codes' new ability to deal with this stiffness and show consistency. 
Then, still in one spatial dimension, we compute a
luminosity versus accretion rate diagram for the setup of
spherical accretion onto a Schwarzschild black hole
and find good agreement with 
previous work which included more radiation processes
than we currently have available. 
 Lastly, we revisit the supersonic \textit{\textbf{B}ondi\textbf{ H}oyle \textbf{L}yttleton} (BHL) accretion in two dimensions where we can now present
simulations of realistic temperatures, down to $T\sim
10^6$\,K or less. 
Here we find that
radiation pressure plays an important role, but also that these highly dynamical set-ups push our approximate treatment towards the limit of
physical applicability.  The main features of radiation
hydrodynamics BHL flows manifest as 
(i) an effective adiabatic index approaching $\gamma_{\rm eff}\sim 4/3$;
 (ii) accretion rates
 two orders of magnitude lower than without radiation
 pressure, but still super-Eddington;
(iii) luminosity estimates around the Eddington limit,
 hence with an overall radiative efficiency as small as
 $\eta_{_{\cal BHL}}\sim 10^{-2}$;
(iv) strong departures from thermal equilibrium in
 shocked regions;
(v) no appearance of the flip-flop  instability.
We conclude that the current optically thick
approximation to the radiation transfer  
does give physically substantial improvements over the pure hydro also
in set-ups departing from equilibrium, and, once
accompanied by an optically thin treatment, is likely to
provide a fundamental tool for investigating accretion
flows in a large variety of astrophysical systems.

\end{abstract}


\section{Introduction}
The field of numerical relativistic hydrodynamics has recently seen 
much progress in treating astrophysical systems
under more and more realistic conditions.
Because of the large computational costs involved, 
the inclusion of multi-dimensional
\textit{\textbf{g}eneral \textbf{r}elativistic
  \textbf{r}adiation\textbf{ h}ydro\textbf{d}ynamics}
(GR-RHD) has been postponed for a long time, with the
remarkable exception  of neutrino transport in the
context of supernovae simulations [see
\citet{Lentz2012} and references therein]. 
However, due to the increasing power of supercomputers, the situation 
has started changing significantly in the last few years, and the
inclusion of a photon-field is no longer regarded as a remote
possibility.

This delay has, however, not been due to the fact that dynamical radiation fields are not regarded as a main ingredient, rather
it is the inherent difficulty of solving the radiation transfer equation\footnote{
See
 \cite{pomraning_1973_erh} and \cite{Mihalas84} for a
 comprehensive treatment of radiation hydrodynamics and  
\cite{Schweizer1988} for the extension to the relativistic case.
}.
The cooling time-scales of a dynamical fluid  may easily
vary over several orders of magnitude within the
computational domain. This then leads to 
characteristic propagation speeds  for the photons
in optically thin regions that are much higher than the
coupled fluid/photon speeds in optically thick regions. 
Not only are time-scales vastly different,
but also additional spatial resolution is required whenever the coupling to the photon field induces  
 small scale instabilities and turbulence.
In addition, surfaces of astrophysical structures are typically 
not in \textit{\textbf{l}ocal \textbf{t}hermal
  \textbf{e}quilibrium} (LTE) 
and can cool very efficiently, usually on much shorter
time scales than the dynamical ones.
This problem becomes particularly severe when performing
global simulations of astrophysical systems in which the principal force is
gravity.
In these cases, the spatial domain must  firstly be large enough
to contain the entire
astrophysical structure and secondly, it needs to resolve the influence of gravity\footnote{
A complementary approach, which is not covered here, is to model not a global 
system, but only a small, representative region e.g. a
shearing box.}. 

Any such multi-scale problem is numerically extremely
costly and it is thus important to formulate efficient
algorithms that include at least a leading order
approximation to the various physics while still
remaining computationally affordable.  
One of the most successful strategies was, and still is,
represented by the so called projected symmetric
trace-free (PSTF) moment formalism introduced by \cite{Thorne1981}.
By defining moments of the
radiation field similarly to how density, momentum and pressure of a
fluid are defined as velocity moments of the corresponding distribution function,
such a formalism provides an accurate, though still reasonably cheap,
approximation to the solution of the radiation transfer
equations. This approach is particularly appealing in the
case of an optically thick medium, characterized by a
strong coupling between matter and radiation. 
\citet{Farris08} were first to undertake the implementation of the
corresponding radiation hydrodynamics equations in a
general relativistic framework. A further step has been
taken by \cite{Shibata2011}, who adopted the variable
Eddington factor approach of \citet{Levermore1984}
to solve
the relativistic radiation-hydrodynamics
equations both in the optically thin and in the optically thick
limit. This represents a significant progress with
respect to simplified treatments, where 
effective cooling functions are introduced.

In spite of all 
this progress, major numerical difficulties 
still prevent the application of such schemes
to realistic astrophysical systems; 
one of them
being the presence of stiff source terms.
For example, in \citet{Zanotti2011} (hereafter paper{\sc I}),
after implementing and testing the framework
suggested by \citet{Farris08}, we
studied the \textit{\textbf{B}ondi   \textbf{H}oyle
  \textbf{L}yttleton} (BHL) accretion flow onto a black
hole, but we could only treat unrealistically high fluid
temperatures of the order of $\sim 10^9$\,K or above.
Though simplified, the BHL
flow can effectively help our understanding of those compact
sources accreting matter with a reduced amount of angular
momentum, and is currently applied to the study of both High
Mass X-ray Binaries \citep{Hadrava2012} and of the
merging of supermassive black hole binaries [see
  \citet{Pfeiffer2012} and references therein]. 

In this paper, we address the problem of treating the
optically thick regime compatible with the conservative
formulation  used in Eulerian GR-MHD codes, while at the
same time coping with the stiffness of the source terms.
As a stiff solver, we choose the
\textit{\textbf{im}plicit-\textbf{ex}plicit} (IMEX)
scheme by \citet{pareschi_2005_ier}, implement  
it in both {\sc Whisky}\footnote{\texttt{www.whiskycode.org}} and
{\sc Echo}\footnote{\citet{DelZanna2007}},
and test the codes against each other. 
As the two codes contain internal differences, such as
scheduling and general infrastructure, it is very useful
to validate them both at this stage, even though the main
part of the simulations 
shown in this paper are performed with {\sc Echo},
because of its spherical, non-uniform grid\footnote{{\sc
    Whisky} uses Cartesian 
adaptive mesh refinement, which is less suited for
spherical models.}. 
  
The paper is organized as follows. In
Sec.~\ref{Radiation_hydrodynamics_in_the_stiff_regime} we
describe the treatment 
of the radiation stiff source terms. We detail an IMEX Runge Kutta
scheme as our 
time integration stiff-solver. Sec.~\ref{Verification_of_the_scheme} presents
the verification of our new scheme through a selected
sample of stiff shock tube problems.
Turning towards astrophysical applications,  we first present in 
Sec.~\ref{sec:spherical} the results for spherical
accretion in a regime that was constructed to be
particularly challenging for the numerics. 
We also present a physical Michel solution and compare it with
previous 
results. Abandoning spherical symmetry, we devote
Sec.~\ref{sec:bh} to the
study of the radiation hydrodynamics of BHL
accretion in two dimensions.
Finally, in  Sec.~\ref{sec:conclusions} we offer a brief summary and our
conclusions. 

Throughout the paper, we set the speed of light $c=1$,
and the gravitational constant $G$ to a pure number.
We extend the geometric units by setting $m_p/k_B=1$,
where $m_p$ is the mass of the proton, while $k_B$ is the Boltzmann
constant. 
However, we have maintained $c$, $G$, and $k_B$ in a explicit form in
those expressions of particular physical interest. 
We refer the interested reader to Appendix~\ref{appendixA} of paper{\sc I} for
the system of extended geometrized units.

\section{Radiation hydrodynamics in the stiff regime}
\label{Radiation_hydrodynamics_in_the_stiff_regime}

\subsection{Formulation of the GR-RHD equations}

In this section, we first review the set of equations
that we use to 
approximate general relativistic 
radiation--hydrodynamics  in the diffusion
limit, as derived in \cite{Farris08} and already
implemented and verified in paper{\sc I}.   
The properties of the fluid immersed in the radiation
field are described by the momentum-energy tensor, which
is given  by
\be
T^{\alpha\beta}=T^{\alpha\beta}_{{\rm m}}+T_{\rm r}^{\alpha\beta}\,,
\ee
%
and comprises a matter contribution
\bea
T^{\alpha\beta}_{{\rm m}}=\rho h\,u^{\,\alpha}u^{\beta}+Pg^{\,\alpha\beta}\,,
\label{eq:T_matter}
\eea
and a radiation contribution
\bea
T_{\rm r}^{\alpha\beta}=\frac{1}{c}\int I_\nu N^\alpha N^\beta d\nu d\Omega \,,
\label{eq:T_radiation}
\eea
where $g^{\alpha\beta}$ is the metric of the spacetime, $u^\alpha$ is
the four-velocity of the fluid, $\rho$, $h=1+\epsilon + P/\rho$,
$\epsilon$, and $P$ are the rest-mass density, the specific enthalpy,
the specific internal energy, and the thermal pressure,
respectively, while
$I_\nu=I_\nu(x^\alpha,N^i,\nu)$ is the specific
intensity of the radiation. We note that $N^\alpha$
defines propagation direction of the photon with
frequency $\nu$, while 
$d\Omega$ is the infinitesimal solid angle around $N^\alpha$.
All of these quantities are measured in the comoving
frame of the fluid. 
The thermal pressure is related to $\rho$ and
$\epsilon$ through an \textit{\textbf{e}quation \textbf{o}f \textbf{s}tate} (EoS), which we take to be that of the
ideal-gas, with constant adiabatic index $\gamma$, i.e.
%
\be
P=\rho\epsilon(\gamma-1) \,.
\ee
In terms of the moments of the radiation field~\citep{Thorne1981}, 
the radiation energy-momentum
tensor $T_{\rm r}^{\alpha\beta}$ can be rewritten as \citep{Hsieh1976}
\be
T_{\rm r}^{\alpha\beta}=(E_{\rm r}+{\cal P}_{\rm r}) u^\alpha u^\beta +
F_{\rm r}^\alpha u^\beta+ u^\alpha F_{\rm r}^\beta + {\cal P}_{\rm r}g^{\alpha\beta} \,,
\ee
where $E_{\rm r}$ and ${\cal P}_{\rm r}$ are the radiation energy
density and pressure, respectively. We make the
additional assumption that the radiation field is 
approximately isotropic, in the sense that ${\cal P}_{\rm r}=E_{\rm
  r}/3$, while the radiation flux is not constrained to
zero, but is allowed to 
take small values such that
$F_{\rm r}^i/E_{\rm r}\ll 1$.
Thus the equations governing the evolution of the system are:
\bea
\label{0eq:mass}
&&\nabla_{\alpha} (\rho u^{\,\alpha})=0, \\
\label{0eq:momentum}
&&\nabla_{\alpha}T^{\alpha\beta}\,\,\,\,\,=0, \\
\label{0eq:rad}
&&\nabla_\alpha T_{\rm r}^{\alpha\beta}\,\,\,\,\,=-G^\beta_{\rm r} \,,
\eea
where $G^\alpha_{\rm r}= G^\alpha_{\rm r}(I,\chi^t ,\chi^s)$, called the radiation four-force
density, depends on the specific intensity and 
on the opacities of the matter interaction.
As in paper{\sc I}, we drop all frequency dependencies
and allow for small deviations from LTE.
We consider bremsstrahlung  and Thomson scattering (i.e. $\chi^t$ and $\chi^s$)
as processes of absorption and scattering.
Using the Planck function, $\tilde{B}$,
it is then possible to write
the radiation four-force in covariant
form as~\citep{Farris08}
\be
\label{four-force}
G^\alpha_{\rm r}=\chi^t(E_{\rm r}-4\pi \tilde{B})u^\alpha+
(\chi^t+\chi^s)F_{\rm r}^\alpha\,.
\ee
In Eq.~(\ref{four-force}) we have introduced the equilibrium
black-body intensity
$4\pi \tilde{B}=a_{\rm rad}T_{\rm fluid}^4$, where $T_{\rm fluid}$
is the temperature of the fluid and $a_{\rm rad}$ is
the radiation constant. We estimate the temperature from the
ideal-gas EoS via the expression
\be
\label{t_estimate}
T_{\rm fluid}=\frac{m_p}{k_B}\frac{P}{\rho}\,,
\ee
where, $k_B$ is the Boltzmann constant and $m_p$ the
rest-mass of the proton. We stress that the method allows
for deviations from thermal equilibrium, namely with $E_{\rm
  r}\neq 4\pi \tilde{B}$.
As shown in paper{\sc I}, after adopting the  
$3+1$ split of spacetime~\citep{Arnowitt62} 
the GR-RHD
equations can be written in conservative form as
\be
\partial_t\bm{\mathcal{U}} + \partial_i\bm{\mathcal{F}}^i=\bm{\mathcal{S}}\,,
\label{eq:UFS}
\ee
where the vector of conserved variables
$\bm{\mathcal{U}}$ and the fluxes $\bm{\mathcal{F}}^i$
are given by
\be
{\bm{\mathcal{U}}}\equiv\sqrt{\text{\textbabygamma}}\left[\begin{array}{c}
D \\ \\ S_j \\ \\U \\ \\ U_{\rm
  r}  \\ \\ (S_{\rm r})_{j}
\end{array}\right],~~~
\bm{\mathcal{F}}^i\equiv\sqrt{\text{\textbabygamma}}\left[\begin{array}{c}
\alpha v^i D-\beta^i D \\\\
\alpha W^i_{\ j}-\beta^i S_j \\\\
\alpha S^i-\beta^i U \\\\
\alpha S^i_{\rm r}-\beta^i U_{\rm r} \\\\
\alpha (R_{\rm r})^i_j-\beta^i (S_{\rm r})_{j} \\\\
\end{array}\right] ,
\label{eq:fluxes}
\ee
while the sources are
\be
\label{eq:sources}
\bm{\mathcal{S}} \equiv \sqrt{\text{\textbabygamma}}\left[\begin{array}{c}
0 \\  \\
\frac{1}{2}\alpha W^{ik}\partial_j\text{\textbabygamma}_{ik}+
S_i\partial_j\beta^i-U\partial_j\alpha+\alpha (G_{\rm r})_j \\ \\
\frac{1}{2}W^{ik}\beta^j\partial_j\text{\textbabygamma}_{ik}+{W_i}^j\partial_j\beta^i
-S^j\partial_j\alpha +\alpha^2 G^t_{\rm r} \\ \\
\frac{1}{2}R^{ik}_{\rm r}\beta^j\partial_j\text{\textbabygamma}_{ik}+(R_{\rm r})_{\ i}^j\partial_j\beta^i
-S_{\rm r}^j\partial_j\alpha-\alpha^2 G^t_{\rm r}\\ \\
\frac{1}{2}\alpha R^{ik}_{\rm r}\partial_j\text{\textbabygamma}_{ik}+
(S_{\rm r})_{i}\partial_j\beta^i-U_{\rm r}\partial_j\alpha-\alpha (G_{\rm r})_j 
\end{array}\right]\,.
\ee
We note that $\alpha$, $\beta$, and $\text{\textbabygamma}$ are the
lapse, the shift, and the determinant of the spatial
metric, respectively,
while $v^i$ and $\Gamma$ are the three-velocity and the
Lorentz factor of the fluid with respect to the Eulerian observer.
In 
the Eqs.~(\ref{eq:fluxes}) and (\ref{eq:sources}) several
more terms have been defined, which we report below for
completeness (c.f. paper{\sc I} for more details):
\bea
W^{ij} & \equiv & \rho h \Gamma^2v^i\,v^j +P\,\text{\textbabygamma}^{ij}\,,
\label{eq:W} \\
S^i & \equiv & \rho h \Gamma^2v^i ,
\label{eq:S} \\
U & \equiv & \rho h \Gamma^2 - P\,,
\label{eq:U} \\
R^{ij}_{\rm r}&=&\frac{4}{3}E_{\rm r} \Gamma^2 v^iv^j
+\Gamma(f_{\rm r}^iv^j+f_{\rm r}^jv^i)+{\cal P}_{\rm r}\text{\textbabygamma}^{ij} \,,
\label{eq:Rperp} \\
S_{\rm r}^i & = & \frac{4}{3}E_{\rm r} \Gamma^2v^i + \Gamma(\alpha
F_{\rm r}^t v^i + f_{\rm r}^i) \,,
\label{eq:S_r} \\
U_{\rm r}&=&\frac{4}{3}E_{\rm r} \Gamma^2 + 2\alpha
\Gamma F_{\rm r}^t -\frac{E_{\rm r}}{3} \,,
\label{eq:U_r} \\
F_{\rm r}^t&=&\frac{v_i F_{\rm r}^i}{\alpha-\beta_i v^i} =
\frac{v_i f_{\rm r}^i}{\alpha}\,.
\eea
%

\subsection{Description of the IMEX scheme for radiation hydrodynamics}

\subsubsection{General concepts}
\label{General_concepts}
A relevant feature of the radiation hydrodynamics
equations \eqref{eq:UFS} is that they contain sources
for the radiation field that may easily become stiff,
depending on the physical conditions under consideration. 
When stiffness is treated
by resorting to
\textit{\textbf{im}plicit-\textbf{ex}plicit} (IMEX)
Runge-Kutta (RK) schemes\footnote{An alternative approach to
solve the special relativistic
RHD equations in a moderately stiff regime
has been considered in one-dimensional
Lagrangian simulations by \citet{Dumbser2012}.},
it is 
important to
split the conservative variables $\bm{\mathcal{U}}$
in two subsets $\{\bm{X},\bm{Y}\}$, with $\{\bm
X\}$ containing the 
variables that are affected by stiffness, and $\{\bm
Y\}$ containing those that are not.
IMEX Runge-Kutta methods are
based on an implicit discretisation for the stiff terms
and on an explicit one for the non-stiff terms.
They have been extensively discussed in a series
of papers by  \citet{pareschi_2005_ier}, and some
recent applications have been presented in special relativistic
 resistive MHD by \citet{Palenzuela:2008sf}, in
general relativistic force-free electrodynamics by \citet{Alic:2012}
and in general relativistic resistive MHD by 
\citet{Bucciantini2012} and by \citet{Dionysopoulou:2012}.
In full generality, the hyperbolic equations for the two sets of
variables $\{\bm{X},\bm{Y}\}$ are split as
\bea
\label{partial_t_Y}
\partial_t \bm{{Y}}&=&F_{\bm{Y}}(\bm{X},\bm{Y})\,,\\
\label{partial_t_X}
\partial_t \bm{{X}}&=&F_{\bm{X}}(\bm{X},\bm{Y})+R_{\bm{X}}(\bm{X},\bm{Y})\,,
\eea
where the operator $F_{\bm{Y}}$ contains both the first
spatial derivatives of $\bm{Y}$ and non-stiff
source terms, the operator  $F_{\bm{X}}$ contains both the first
spatial derivatives of $\bm{X}$ and non-stiff
source terms, while the operator $R_{\bm{X}}$ contains
the stiff source terms affecting the variables $\bm{X}$.
Each Runge-Kutta sub-stage of the IMEX scheme can be divided
in two parts.
\begin{enumerate}
\item
In the first part, 
the explicit intermediate values $\{\bm{X}^{\ast,i},\bm{Y}^{\ast,i}\}$ 
of each sub-stage $i$ are computed as
\bea
\bm{Y}^{\ast,i}&=&\bm{Y}^n+\Delta
t\sum_{j=1}^{i-1}\tilde{a}_{ij}F_{\bm{Y}}[\bm{\mathcal{U}}^{(j)}]\,,\\
\bm{X}^{\ast,i}&=&\bm{X}^n+\Delta
t\sum_{j=1}^{i-1}\tilde{a}_{ij}F_{\bm{X}}[
  \bm{\mathcal{U}}^{(j)}]+\Delta t
\sum_{j=1}^{i-1}{a}_{ij}R_{\bm{X}}[\bm{\mathcal{U}}^{(j)}]\,,\nonumber \\
\eea
where one might note that the summation stops at
$(i-1)$, in order to avoid the appearance of the implicit
terms at this stage. The matrices $(\tilde
a_{ij})$ and $(a_{ij})$ are $\nu\times\nu$ square
matrices.  In this paper, we use $\nu=4$
(see also Appendix \ref{appendixB}),
whereas, in general, the matrix coefficients
and dimensions change with the desired 
number of stages\footnote{Note that the  global order of
  an IMEX scheme does not uniquely determine the number of sub-stages.}
\citep{pareschi_2005_ier}. 
\item
In the second part, the non-stiff
variables are directly advanced to the status of sub-stage
Runge-Kutta variables, namely
\be
\bm Y^{(i)}=\bm Y^{\ast,i}\,,
\ee
while the stiff variables need to be corrected as
\be
\label{correct_X}
\bm X^{(i)}=M(\bm{Y}^{\ast,i}) \left[ \bm{X}^{\ast,i}
  +a_{ii}\Delta t \bm{K}_{\bm X}(\bm{Y}^{\ast,i}) \right] \,.
\ee
The vector $K_{\bm X}(\bm{Y})$ on the right hand side of
Eq.~\eqref{correct_X}, which does not depend on the stiff
variables $\bm X$,
results from the decomposition of $R_{\bm X}(\bm{X},\bm{Y})$
as 
\be
\label{R_dec}
R_{\bm X}(\bm{X},\bm{Y})=A(\bm{Y})\bm X + K_{\bm X}(\bm{Y})\,,
\ee
while the matrix $M$ is given
by~\citep{Palenzuela:2008sf}\footnote{We stress that
the form of $M$ given by Eq.~\eqref{matrix-M}
is only valid for the decomposition as done in Eq.~\eqref{R_dec}.}
\be
\label{matrix-M}
M(\bm{Y}^{\ast,i})=\left[I-a_{ii}\Delta t A(\bm Y^{\ast,i})\right]^{-1}\,,
\ee
where $I$ is the identity matrix.
\end{enumerate}
%
%
%
%
For each RK sub-stage,  $\{\bm{X}^{(i)},\bm{Y}^{(i)}\}$ is computed
as described above, and finally the time-update is performed as
\be
\bm{\mathcal U}^{n+1}=\bm{\mathcal U}^{n}+\Delta t
\sum_{i=1}^\nu \tilde w_i F[\bm{\mathcal U}^{(i)}]+
\Delta t
\sum_{i=1}^\nu w_i R[\bm{\mathcal U}^{(i)}]\,,
\ee
where $\tilde w_i$ and $\omega_i$ are coefficient
vectors. In most of the applications presented in this
paper, we have adopted the SSP3$(4,3,3)$ (Strong Stability
Preserving of order three) IMEX Runge-Kutta
scheme. 
The notation SSPk$(s,\sigma,p)$ is adopted
to specify the order of the SSP scheme ($k$), the
number of 
stages of the implicit scheme ($s$),  the
number of 
stages of the explicit scheme ($\sigma$),  and
the order of the IMEX scheme ($p$)~\citep{pareschi_2005_ier}. 
The coefficient tables employed in this paper are listed
in the Appendix \ref{appendixB}.
%

\subsubsection{Specification to radiation hydrodynamics}
\label{Specification_to_radiation_hydrodynamics}

Because of the complexity of the GR-RHD 
equations, isolating the
term (or the terms) that are responsible for the
stiffness is not a trivial task, although we can
certainly say that such terms are contained in the
radiation four-force $G_{\rm r}^\alpha$.
According to the logic of the IMEX scheme just described,
we identify $\{\bm X\}$ with the radiation hydrodynamical
variables $\{U_{\rm r},(S_{\rm r})_j\}$
that are affected by stiffness, and $\{\bm Y\}$ with $\{D,S_j,U\}$, 
that remain unaffected.

As highlighted above, the IMEX scheme requires the
stiff source terms $R_{\bm X}$ to be decomposed according to
Eq.~\eqref{R_dec}. We therefore 
write the radiation four-force $G_{\rm r}^\alpha$ in terms of the
conservative variables of the radiation field. 
To this extent, we rewrite Eq.~\eqref{eq:S_r} and Eq.~\eqref{eq:U_r} to find
the radiation energy density $E_{\rm r}$ and the fluxes $F^\alpha_{\rm
  r}$ in terms of $U_{\rm r}$ and $(S_{\rm r})_i$ as
\bea
E_{\rm r} &=& -3 \Gamma^2 W \left[2 (S_{\rm r})_kv^k +
U_{\rm r} (1/\Gamma^2-2)\right]\,,\\
F_{\rm r}^t &=& \frac{\Gamma}{\alpha} W \left[ -4 U_{\rm
    r} (\Gamma^2 -1) + (4 \Gamma^2 -1) (S_{\rm r})_kv^k\right]\,,\\
(f_{\rm r})_i &=& \frac{(S_{\rm r})_i}{\Gamma} -
\frac{4}{3} E_{\rm r} \Gamma v_i - \alpha (F_{\rm r})^t v_i\,,
\eea
where $W=1/(1+2\Gamma^2)$. In this way, and after some
simple algebra, we can rewrite the
radiation four force as
\begin{widetext}
\bea
\label{Gt_cons}
G_{\rm r}^t = &-& \frac{\Gamma}{\alpha} \left[ \chi^t a_r T_{\rm fluid}^4+
  U_{\rm r} (2 \chi^s(1-3W)-\chi^t) + (S_{\rm r})_k v^k
  (\chi^t + \chi^s(3W-2)) \right]\,,\\
\label{Gi_cons}
(G_{\rm r})_i = &-& \chi^t a_r T_{\rm fluid}^4 v_i\Gamma +\frac{(\chi^t +
  \chi^s)}{\Gamma}(S_{\rm r})_i +U_{\rm r} \Gamma v_i \left[\chi^t(1-4W)+2 \chi^s(W-1) \right]
      +  (S_{\rm r})_kv^k  \Gamma v_i \left[ \chi^t(2W-1)+\chi^s (2-W) \right]\,.\nonumber\\
\eea
\end{widetext}
We note that the right hand sides of \eqref{Gt_cons}
and \eqref{Gi_cons} 
do not contain the set of variables $\{\bm Y\}$, while
they do contain the conserved variables $\{\bm X\}$,
which always appear with a
multiplication factor containing either $\chi^t$ or
$\chi^s$. This is an indication that, depending on the
values assumed by the opacities, such source terms may
become stiff, {\em but only for the radiation variables}. 
This means that the vector of sources given by Eq.~\eqref{eq:sources}
will be split in two parts,
$\bm{\mathcal{S}}=\bm{\mathcal{S}}_e+\bm{\mathcal{S}}_i$. The
first one,
\be
\label{eq:sources_e}
\bm{\mathcal{S}}_e \equiv \sqrt{\text{\textbabygamma}}\left[\begin{array}{c}
0 \\  \\
\frac{1}{2}\alpha W^{ik}\partial_j\text{\textbabygamma}_{ik}+
S_i\partial_j\beta^i-U\partial_j\alpha+\alpha (G_{\rm r})_j \\ \\
\frac{1}{2}W^{ik}\beta^j\partial_j\text{\textbabygamma}_{ik}+{W_i}^j\partial_j\beta^i
-S^j\partial_j\alpha +\alpha^2 G^t_{\rm r} \\ \\
\frac{1}{2}R^{ik}_{\rm r}\beta^j\partial_j\text{\textbabygamma}_{ik}+(R_{\rm r})_{\ i}^j\partial_j\beta^i
-S_{\rm r}^j\partial_j\alpha\\ \\ 
\frac{1}{2}\alpha R^{ik}_{\rm r}\partial_j\text{\textbabygamma}_{ik}+
(S_{\rm r})_{i}\partial_j\beta^i-U_{\rm r}\partial_j\alpha 
\end{array}\right]\,,
\ee
will be absorbed into the operators $F_{\bm Y}$ and
$F_{\bm X}$ in Eqs.~\eqref{partial_t_Y} and~\eqref{partial_t_X},
because it 
does not contain stiff terms. The second part, on the other hand, 
which contains the genuinely stiff terms for the
radiation variables $\{\bm X\}$, is
\be
\label{eq:sources_i}
\bm{\mathcal{S}}_i \equiv \sqrt{\text{\textbabygamma}}\left[\begin{array}{c}
0 \\  \\
0 \\ \\
0 \\ \\
-\alpha^2 G_{\rm r}^t \\  \\
-\alpha (G_{\rm r})_j 
\end{array}\right],
\ee
and its non-zero components are identified with $R_{\bm
  X}(\bm X,\bm Y)$ in Eq.~\eqref{partial_t_X}.
After using Eq.~\eqref{Gt_cons} and
Eq.~\eqref{Gi_cons}, it is possible to further decompose $R_{\bm
  X}$ as prescribed by Eq.~\eqref{R_dec} as\\
%
\be
\label{explicit_dec}
\left[\begin{array}{c}
-\alpha^2 G_{\rm r}^t \\ \\-\alpha (G_{\rm r})_j
\end{array}\right]=~~~ A(\bm Y) ~~~
\left[\begin{array}{c}
 U_{\rm r}  \\ \\ (S_{\rm r})_j
\end{array}\right]+
\left[\begin{array}{c}
\alpha\Gamma\chi^t a_r T_{\rm fluid}^4  \\ \\ \alpha\Gamma\chi^t a_r T_{\rm fluid}^4v_j
\end{array}\right]\,,
\ee
where the coefficients of the matrix $A(\bm Y)$ are specified in the Appendix
\ref{appendixB}. The components of the vector $K_{\bm X}$ (the
second term on the right hand side of
Eq.~\eqref{explicit_dec}) do not depend on the stiff
variables $\bm X$, but only on the temperature $T_{\rm fluid}$.
We note that, in the actual implementation of the IMEX Runge-Kutta
scheme, the correction to the implicit variables $\bm
X^{(i)}$ dictated by Eq.~\eqref{correct_X}
is performed when the conversion from the conservative
variables $\bm{\mathcal U}$
to the primitive variables is performed.\\

\subsection{Numerical tools}
For reasons of flexibility, cross-verification and in view of future projects,
we have implemented the GR-RHD equations in their IMEX version
in two different numerical codes. 

The first one is a modification of the {\sc Whisky} code,
which implements the general relativistic resistive
magnetohydrodynamics formalism 
{\sc WhiskyRMHD} \citep{Dionysopoulou:2012}. We use the numerical methods provided
by the original {\sc Whisky} code documented in \citep{Baiotti03a,Giacomazzo:2007ti}, namely an HLLE approximate
Riemann solver and a second order TVD slope limiter method
 for the reconstruction of the primitives.   
The infrastructure as well as the
solution of the Einstein equations is provided by the
\texttt{Cactus} Computational Toolkit~\citep{Loffler:2011ay}.
The implementation of the GR-RHD equations in {\sc WhiskyRMHD} 
required modifications mainly in the sources and the routine which recovers the primitives from the
conservative variables. 
In order to deal with the stiffness of the source terms,
we have modified the MoL thorn (part of
Einstein Toolkit\footnote{\cite{cactusweb}}), by including second and
third-order IMEX Runge-Kutta time integrators. 

The second code is based on {\sc Echo}
\citep{DelZanna2007}, which provides a numerical platform for
the solution of the GRMHD 
equations in stationary background
spacetimes\footnote{See also \citet{Bucciantini2011} for
  a recent extension of ECHO to dynamical spacetimes
  within the conformally flat approximation.}. 
It employs a high-order shock-capturing Godunov scheme
with a two-waves HLL Riemann solver, while the spatial
reconstruction of the primitive 
variables can be obtained by linear
and non-linear methods. Time integration is possible in either second or third-order
IMEX Runge-Kutta.
Previously in paper{\sc I}, {\sc Echo} had been
extended to allow the solution of
the non-stiff GR-RHD equations in the optically thick regime.

In both {\sc Whisky} and {\sc Echo}, our implementation of the stiff
GR-RHD equations  
does not allow for a treatment of the
optically \textit{thin} regime. 
Therefore, all the tests and applications 
described in this paper are limited to the optically
\textit{thick} regime, while we postpone  an accurate analysis of the variable Eddington
factor approach to a future work.

Finally, we note that the increase of computational cost when changing
  from an explicit RK of order $k$ to
a RK-IMEX of the same order $k$ is approximately given by the ratio of
the number of sub-stages required by the IMEX to
the number of sub-stages required by the explicit RK.
For the SSP3-IMEX scheme, compared to the explicit RK3,
such a nominal ratio is given by
$5/3\sim 1.67$ and in both our implementations we
have measured an effective factor $\sim 1.8$ increase.

%
\begin{table*}
\begin{center}
\caption{{\sc Description of the initial data } -  in the shock-tube tests
  with radiation field. The different columns refer respectively to:
  the test considered,  the adiabatic index, the radiation constant
  and the thermal opacity. Also reported are the rest-mass density,
  pressure, velocity and radiation energy density in the ``left''
  ($L$) and ``right'' ($R$) states.}
\label{tab1}
\begin{tabular}{llllllllllll|}
\hline
\hline
Model & $\gamma$ & $a_{\rm rad}$ & $\kappa_g^t$ & $\rho_L$
& $P_L$ &  $u^x_L$ & $E_{{\rm r},L}$ & $\rho_R$ & $P_R$ &
$u^x_R$ & $E_{{\rm r},R}$    \\

\hline

\texttt{1} & $2$   &$1.543\times10^{-7}$ &$25$ &$1.0$ &$60.0$ & $10.0$ & $2.0$&$8.0$ &
$2.34\times 10^3$  &  $1.25$ & $1.14\times 10^3$ \\
\texttt{2} & $5/3$ &$1.388\times10^8$ &$0.7$ & $1.0$ &
$6.0\times10^{-3}$ & $0.69$ &  $0.18$ & $3.65$ & $3.59\times 10^{-2}$  &  $0.189$ & $1.3$ \\
\texttt{3} & $2$   &$1.543\times10^{-7}$ &$1000$ &$1.0$ &$60.0$ & $1.25$ & $2.0$&$1.0$ &
$60.0$  &  $1.10$ & $2.0$ \\
\hline
\hline
\end{tabular}
\end{center}
\end{table*}
%

\begin{figure*}
{\includegraphics[angle=0,width=8.0cm,height=7.5cm]{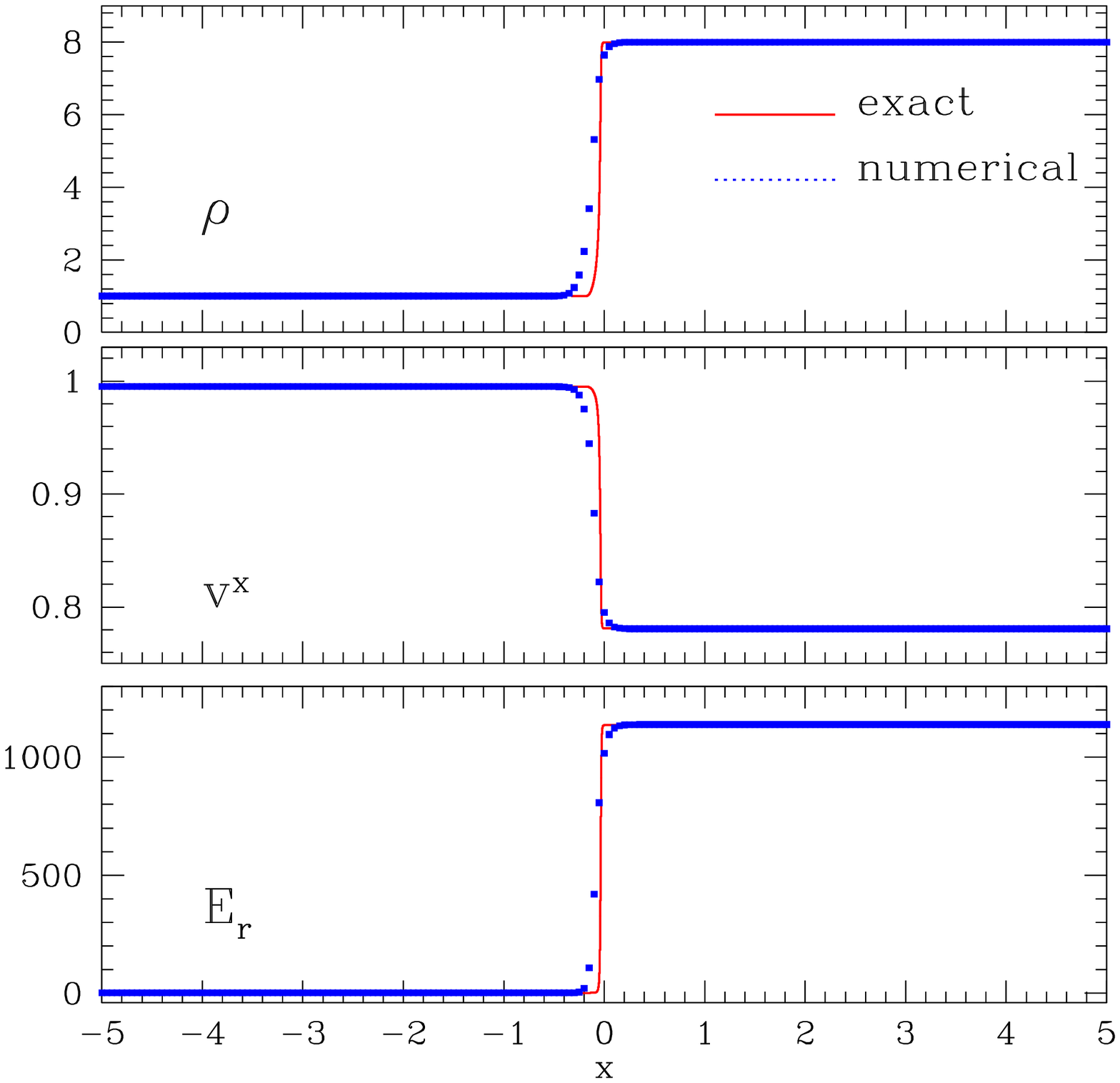}}
\hskip 1.0cm
{\includegraphics[angle=0,width=8.0cm,height=7.5cm]{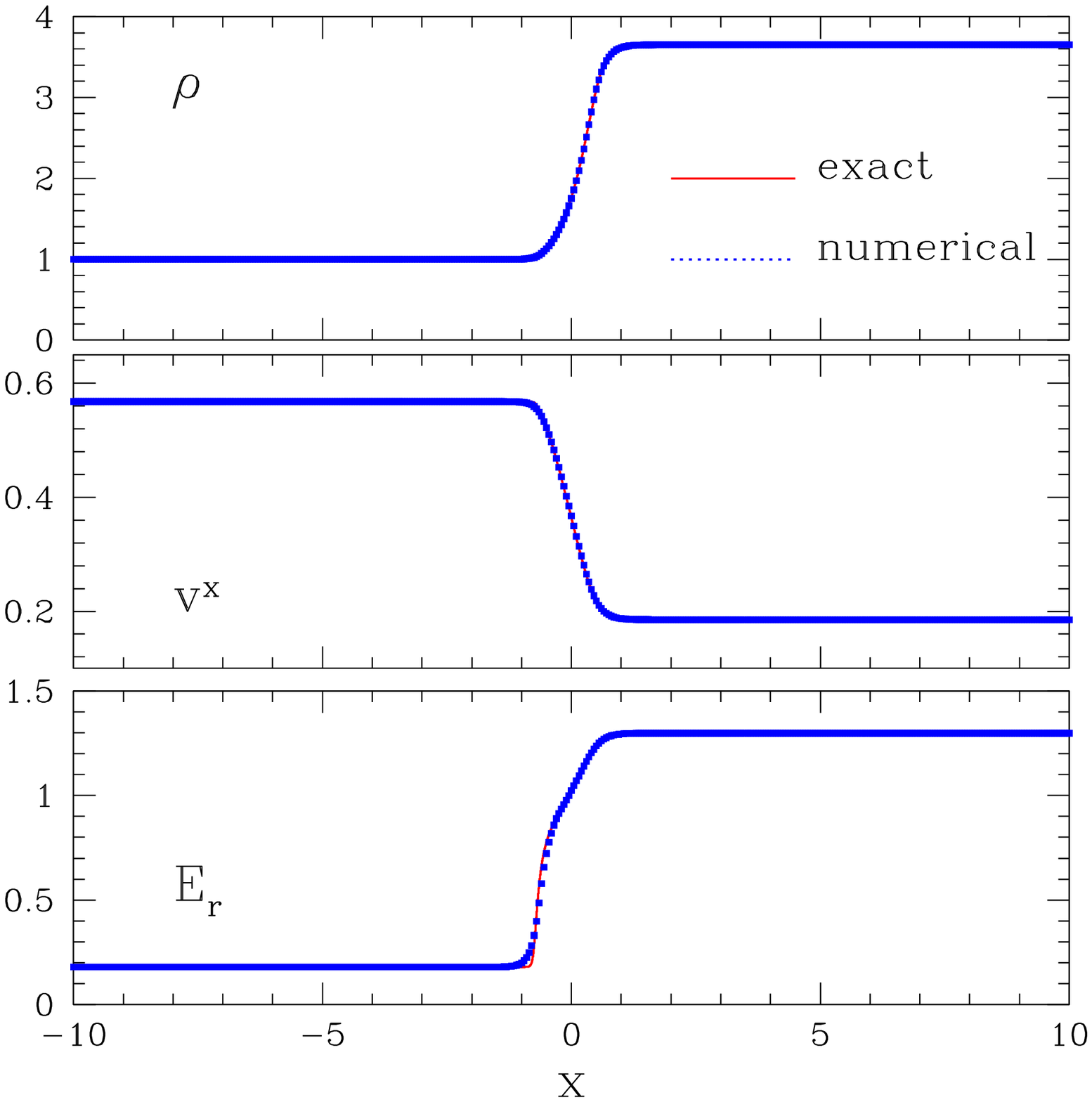}}
\vspace*{-0.0cm}
\caption{{\sc Shock tubes} - Solution of the test No. \texttt{1} 
(left panel) and No. \texttt{2} 
  (right panel). From top
  to bottom the panels report the rest-mass density, the velocity and the
  radiation energy density.
  In both cases $800$ grid-points have been used with
  $\cfl=0.25$ and RKIMEX2. The tests are performed with the
  {\sc Whisky} code, employing TVD reconstruction and minmod limiter.}
\label{fig:test1-2}
\end{figure*}

\section{Verification of the scheme}
\label{Verification_of_the_scheme}

In paper{\sc I}  we had considered a number of
shock-tube tests in which nonlinear
radiation-hydrodynamic waves propagate. 
The semi-analytic solution that is used for comparison with the numerical
one has been obtained following the strategy of~\citet{Farris08}, and
it requires the solution of the following system of ordinary
differential equations
\be
\label{ode}
d_x \mathbf{U}(\mathbf{P}   ) = \mathbf{S}(\mathbf{P}) \,,
\ee
where
\[\mathbf{P} = \left(
\begin{array}{c}
\rho\\
P\\
u^x\\
E_{\rm r}\\
F_{\rm r}^x
\end{array}\right)\,,
\qquad
\mathbf{U} = \left(
\begin{array}{c}
\rho u^x\\
T^{0 x} \\
T^{x x}\\
T_{\rm r}^{0 x} \\
T_{\rm r}^{x x}
\end{array}\right)  \,,
\qquad
\mathbf{S} =\left(
\begin{array}{c}
0\\
0\\
0\\
 -G^0_{\rm r}\\
 -G^x_{\rm r}
\end{array}\right)\,.\]
$U_1$, $U_2$ and $U_3$ are constant in $x$,
while only $T_{\rm r}^{0 x}$ and $T_{\rm r}^{x x}$ need
to be solved for. These tests can be used to monitor the
ability of the code 
to deal with the stiff regime, by
simply increasing the thermal opacity $\kappa_g^t$ 
(the scattering opacity $\kappa_g^s$ is set to zero).
 When this is done, the semi-analytic
solution of the ODE system (\ref{ode}) can be obtained
with an ODE solver for stiff systems~\citep{Press92}.
The initial states of the two
tests that we have considered 
are reported in Table~\ref{tab1} and are chosen in such a way
that the discontinuity front at $x=0$ remains stationary, namely it
is comoving with the Eulerian observer.
LTE
is assumed at both ends $x=\pm X$, with $X=20$, and this is obtained
by adopting a fictitious value of the radiation constant $a_{\rm
  rad}$, namely $a_{\rm rad}=E_{{\rm r},L}/T_L^4$, which is then used
to compute $E_{{\rm r},R}=a_{\rm rad}T_R^4$ (here the indices $L$ and
$R$ indicate the ``left'' and ``right'' states,
respectively). 
We note that tests No. \texttt{1} and \texttt{2} in
Table~\ref{tab1} are the same of 
tests No. \texttt{3} and \texttt{4} in
Table~\ref{tab1} of \citet{Zanotti2011}, apart for the
value of $\kappa_g^t$, which controls the stiffness of
the problem. After setting $800$ grid-points in the $x$ direction,
we have increased the value of $\kappa_g^t$
to the maximum value affordable by the numerical scheme, 
{\em while keeping the \cfl\, parameter unchanged
  and equal to $0.25$.} For example, $\kappa_g^t$ has been
increased from $0.3$ to $25.0$ in test No. \texttt{1}, and 
from $0.08$ to $0.7$ in test No. \texttt{2}.
Each test is evolved in time until stationarity is reached, and the results
are shown in  Figure~\ref{fig:test1-2}, where the numerical
solution is compared to the semi-analytic one in the
two cases considered.

\begin{figure}
{\includegraphics[angle=0,width=8.0cm,height=7.5cm]{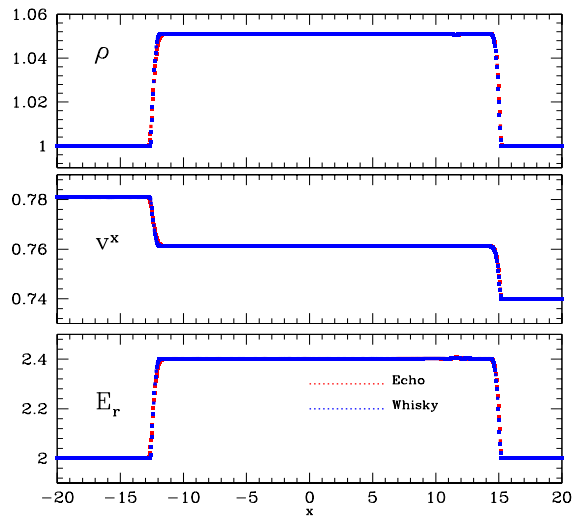}}
\vspace*{-0.0cm}
\caption{{\sc Shock tubes} - Solution of the test No. \texttt{3} at
  time $t=15$.
  From top
  to bottom the panels report the rest-mass density, the velocity and the
  radiation energy density.
  In both cases $800$ grid-points have been used with
  $\cfl=0.25$ and RKIMEX2. 
  The tests are performed with both the
  {\sc Whisky} and {\sc Echo} codes, employing TVD reconstruction and MC limiter.
}
\label{fig:test3}
\end{figure}

It should also be noted that shock tube problems do not represent an ideal
set-up to highlight the ability of the scheme in handling 
the stiffness of the source terms, since strong discontinuities 
are by themselves a challenge for any numerical method. As a result, we have
performed an additional and  
peculiar shock tube problem, test No. \texttt{3}, which  
has equal left and right states, except for the velocity. In this case, 
two shock waves propagate in opposite direction, no stationary solution is 
obtained, but a much higher value of $\kappa_g^t$ can be used, namely 
$\kappa_g^t=1000$. 
Figure~\ref{fig:test3} reports the corresponding solution at time $t=15$,
and also shows the very good agreement between the results obtained with 
{\sc Whisky} and {\sc Echo}.
 
\section{Spherical accretion}
\label{sec:spherical}

Having introduced the numerical tools for the treatment
of the stiff source terms typical of GR-RHD,
we now focus on a problem that has been the subject of
several astrophysical analyses, namely spherical accretion
onto a black hole.
In the first part of this \S, we 
present an additional 
test of our numerical scheme, brought in the stiff regime by assuming unphysically large
cross-sections.  
On the other hand,  in the second part, we choose physical
parameters to model the solution by
\citet{Michel72} in an astrophysical context.

Transonic accretion onto a
non-rotating black hole in the presence of an isotropic radiation
field has been studied in great detail by several
research groups over the years. In the optically thick regime, the
stationary solution was investigated under different approximations
and by focusing on different emission mechanisms by
\citet{Maraschi1974}, \citet{Kafka1976}, \citet{Vitello1978},
\citet{Gillman1980}, \citet{Flammang1982}, \citet{Nobili1991}.  The
time dependent solution, 
was considered by
\citet{Gilden1980} and \citet{Zampieri1996}. The latter, in
particular, 
solved via a Lagrangian code the radiation transfer
equations using the PSTF moment
formalism\footnote{The first time dependent problems
  adopting the PSTF formalism were presented in \citet{Rezzolla1994}.}, 
truncated at the first two moment
equations. Because of the limiting approximations assumed, and in
particular because of the lack of Comptonization effects, our analysis
should not be regarded as an attempt to improve with respect to the
above mentioned works, but rather as a preliminary study in view of
further developments. We also note that multidimensional
simulations with an Eulerian code have been recently
performed by \citet{Fragile2012} obtaining promising results.

\begin{figure}
{\includegraphics[angle=0,width=8.0cm,height=7.5cm]{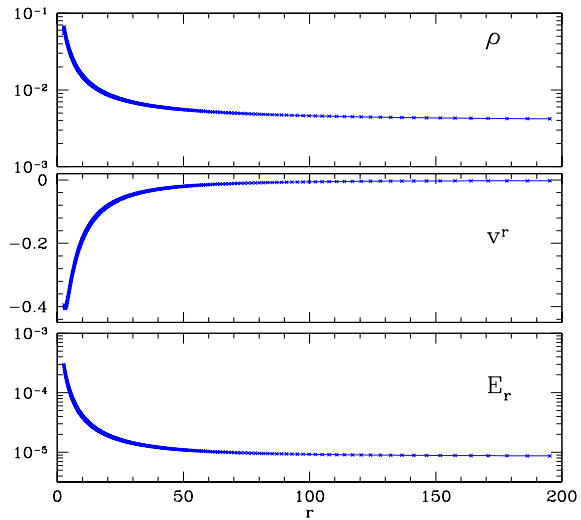}}
\vspace*{-0.0cm}
\caption{{\sc Stiff spherical accretion} - Numerical solution at time $t=1000$.
From top to bottom the panels report the rest-mass
density, the velocity, the radiation energy density.
An artificial $\kappa_g^t=1.0\times10^{15}$ has been
adopted to highlight the ability of the code 
to treat the stiff regime.
$N_r=300$ grid-points have been used with
$\cfl=0.2$, MC reconstruction  and SSP3-IMEX.
}
\label{fig:spherical_accretion}
\end{figure}

Our initial conditions are given by the fluid spherically symmetric
transonic solution of \citet{Michel72}, which is stationary in the
absence of a radiation field. The free parameters of the fluid
solution are the critical radius $r_c$
and the rest mass density at the critical radius
$\rho_c$. 
We choose a black hole with mass $M=2.5 M_\odot$,
while the adiabatic index of the fluid is $\gamma=4/3$. 
The initial radiation field is initialized to a negligible energy
density, while radiation fluxes are set to zero.  
As a first test, aimed at showing the ability of the
numerical scheme in handling the stiff regime, we have
considered an unphysical setup with $\rho_c=0.02$,
$r_c=8.0$, and a high uniform value of the thermal opacity,
$\kappa_g^t=10^{15}$.
The test is performed in
Boyer-Lindquist coordinates with $2.5<r<200$ using $N=300$ radial grid
points. The SSP3-IMEX scheme has been adopted, with the MC limiter for
the spatial
reconstruction. Figure~\ref{fig:spherical_accretion} 
shows the profiles of the rest-mass density, the
radial velocity and the radiation energy density
(from top to bottom) at time $t=1000$.
We stress that, if the IMEX scheme is not available,
and the evolution is performed through a fully explicit
Runge-Kutta scheme, this test can be successfully
repeated at the same $\cfl=0.2$ only with a value of $\kappa_g^t\lsim 1$.

Having done that, we have concentrated on a sequence of
more realistic models, all of them with 
$\rho_c=9.88\times10^{-9}{\rm cgs}$, but with different
critical radii, chosen in the range
between $r_c=800$ and $r_c=7000$,
in order to control the accretion rate.  The test is performed in
Kerr-Schild coordinates with $1.0<r<1000$ using $N=3200$ radial grid
points The evolution is stopped when stationarity in the $L2-$norms of
all of the variables has been reached, which may require a final time
as long as $t=400000$ in code units.  The scheme employed
is the SSP3-IMEX, with MC limiter. 

Special attention
has to be paid to the boundary conditions at the outer radial grid
point, for which we have followed closely the discussion presented
by~\citet{Nobili1991}.  In particular, zeroth order extrapolation
(copy of variables) is adopted for the gas pressure and for the
density. This guarantees that the temperature has zero gradient. At
the same time we want to make sure that at large radii the radiation
field streams radially, namely that $E\propto f^r \propto
r^{-2}$. This translates into the condition
\be
\frac{d\ln E}{d\ln r}=-2 \ ,
\ee
which can be easily implemented. Finally, we fix the accretion rate at
the outer boundary to the value possessed by the initial
configuration.  At the inner radial boundary, on the other hand,
zeroth order extrapolation is adopted for all of the variables.  The
evolution is performed considering both the contribution of the
bremsstrahlung opacity and of the Thomson scattering opacity for
electrons.  We note that during the evolution the radiation flux
remains typically two orders of magnitude smaller than the radiation
energy, thus maintaining the code in the physical regime 
for which it was designed.
After an initial relaxation, the system converges to a
different stationary configuration characterized by a non-zero
radiation flux. The solution is optically thick in all the models
for $r\leq 100$, while it becomes marginally optically
thick at large radii.  From
the radiation flux we compute the luminosity as $L=4 \pi r^2 f^r$.

\begin{figure}
{\includegraphics[angle=0,width=8.0cm,height=7.5cm]{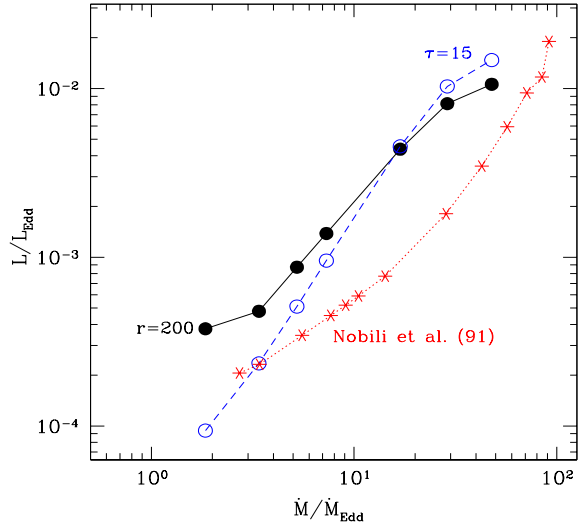}}
\vspace*{-0.0cm}
\caption{{\small{\sc Spherical accretion: luminosity(accretion rate $\dot{M}$)}} - in Eddington
units. Luminosity $L$ was extracted either at constant optical depth $\tau$ or at constant radius $r$.
 Additionally in \textcolor{red}{red}, we show the high luminosity branch found in \citet{Nobili1991}
for comparison.
}
\label{fig:test_michel}
\end{figure}

Fig.~\ref{fig:test_michel} reports the results of our simulation tests
in the diagram $(\dot{M}/\dot{M}_{\rm Edd},L/L_{\rm Edd})$, where the
luminosity is computed at $r=200$.  Although our data resemble the
high luminosity branch reported in Fig. 1 by \citet{Nobili1991}, a
close comparison with their results is not really possible, since
Comptonization, bound-bound transitions and free-bound transitions are
not taken into account in our analysis.  In particular, the absence of
pre-heating effects does not allow us to verify the onset of strong
thermal instabilities producing hydrodynamic shock waves that
propagate outward, as reported by \citet{Zampieri1996}.  In spite of
this, the test we have performed is very relevant.  In fact, by using
an entirely different procedure with respect to \citet{Nobili1991} and
\citet{Zampieri1996}, it confirms the existence of a high luminosity
branch in the diagram $(\dot{M}/\dot{M}_{\rm
  Edd},L/L_{\rm Edd})$, which
corresponds to the optically thick regime.  A more extended analysis
of this problem, by including additional contributions to the opacity,
a treatment of the Comptonization and the effect of a
spinning black hole will be the focus of a separate and dedicated
work.

\section{Bondi--Hoyle--Lyttleton (BHL) accretion}
\label{sec:bh}
This Section deals with the application of our new scheme
to simple astrophysical models departing from spherical symmetry.
We revisit the BHL accretion flow,
that we already described in some detail in paper{\sc I}, and
whose initial conditions are briefly summarized in
Sec.~\ref{sec:Initial_conditions}. 
After showing consistency with the non-stiff solver, 
we illustrate the effectiveness of the IMEX by treating
models of low temperature, which is the key parameter
responsible for numerical difficulty. 
Only now, it becomes feasible to treat
astrophysical temperatures that are few orders of magnitude
lower than in paper{\sc I} and as astrophysically realistic
as our approach can allow at this stage (c.f.
conclusion for more discussion).
Having thus reached 
the limits imposed by the physical assumptions of the current treatment,
we now analyse the dynamics of the fluid, the
occurrence of shocks and the possible observational 
quantities, with particular attention to the computation
of the luminosity. 
We stress that it is
not our intention to perform a systematic analysis of the full
parameter space.
\subsection{Initial conditions for the BHL accretion flow}
\label{sec:Initial_conditions}
We perform two dimensional numerical simulations 
of a BHL accretion flow \citep{Hoyle1939,Bondi1944} onto a Schwarzschild 
black hole of galactic
size with $M_{BH}=3.6\times 10^6\, M_{\odot}$.
The initial conditions considered are similar to those
adopted in paper{\sc I}, with a
velocity field that is specified in terms of
an asymptotic velocity $v_{\infty}$, ~\citep{Font98a}
\begin{eqnarray}
\label{asymtotic_vel_bla}
v^r &=& \sqrt{\text{\textbabygamma}^{rr}} v_{\infty} \cos\phi \,,\\
v^{\phi} &=&  -\sqrt{\text{\textbabygamma}^{\phi \phi}}  v_{\infty}
\sin\phi \,,
\label{asymtotic_vel_blb}
\end{eqnarray}
 where {\textbabygamma}$^{ij}$ are the components of
    the 3-metric and
    $\phi$ is the azimuthal angle in Boyer-Lindquist coordinates.
The radiation field is initialized 
to a uniform and small energy density $E_{\rm r}$, 
such that the
radiation temperature $T_{\rm rad}=(E_{\rm r}/a_{\rm
  rad})^{1/4}\approx 1.5\times 10^5 K$. 
Additional free parameters are
the asymptotic sound speed $c_{s,\infty}$, and the
asymptotic pressure, from which the asymptotic rest-mass density
$\rho_{\infty}$ follows directly.
The resulting configuration relaxes 
to a different and stationary one, on a timescale that depends on the
parameters chosen. Keeping to nomenclature of paper{\sc I},
we encode the two 
parameters $v_{\infty_{0.1}}$ and
$c_{s,\infty_{0.1}}$\footnote{Here, subscripts $0.1$
  denote 
 the normalisation in units 
of $0.1\, c$, so
$v_{\infty_{0.1}}=v_{\infty}/(0.1\,c)$. Therefore, the
model ${\tt V07.cs03}$ has $v_{\infty}=0.07$ and $c_{s,\infty}=0.03$. 
} and a prefix denoting perturbation (if applicable) 
in our naming scheme as ${\tt
  \_\_.v_{\infty_{0.1}}.c_{s,\infty_{0.1}}}$. The
adiabatic index of the fluid is $\gamma=5/3$. \\
The computational
grid consists of $N_r \times N_{\phi}$ numerical cells in the radial
and angular directions, respectively, covering a computational domain
extending from $r_{\rm {min}}=2.1\,M$ to $r_{\rm {max}}=200\,M$ and
from $\phi_{\rm {min}}=0$ to $\phi_{\rm {max}}=2\pi$. For our fiducial
simulation we have chosen $N_r=1536$ and $N_\phi=300$, but have also
verified that the results are not sensitive to the resolution used or
to the location of the outer boundary.
\subsection{Consistency test}
\begin{figure}
\includegraphics[angle=0,width=8.0cm,height=7.5cm]{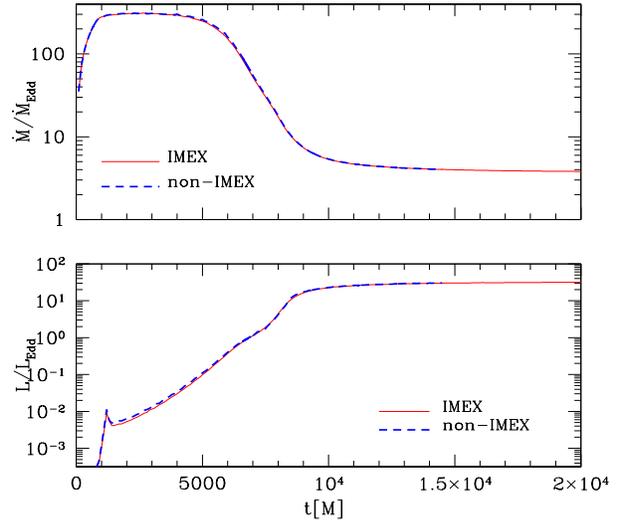}
\caption{{\sc Consistency test of the new IMEX scheme} - Time evolution of a perturbed BHL model with
$v_{\inf}=0.18$ and $c_{s,\infty}=0.07$
(model ${\tt p.V18.cs07}$ of paper{\sc I}) with the \textcolor{red}{IMEX}
(solid line) and with the \textcolor{blue}{non-IMEX} version of
the code (dashed line).
The lower panel shows the light curve, whereas 
the accretion rate is plotted in the top panel. All curves are shown in
Eddington units. 
}
\label{fig:consistency}
\end{figure}

Before going to new models, we first carried 
out a consistency test 
using a representative model with Mach number ${\cal M}_{\infty}=2.57$
(model ${\tt p.V18.cs07}$
of paper{\sc I}) and reproducing it with the present
new IMEX-version of {\sc Echo}. 
As shown in Fig.~\ref{fig:consistency}, 
the IMEX version reproduces the light curves and
the accretion rates obtained with the purely
explicit version of the code. 
Moreover, by using the IMEX scheme, 
it is now possible to extend the evolution to later times,
whereas
the previous version of the code 
required reducing the \cfl\, to values smaller than  
$0.01$, making such long evolutions practically
unfeasible.  
This test confirms
that the new scheme is verified also 
in a non-trivial two-dimensional application and
that the use of the IMEX offers clear advantages in
terms of computational resources.
\subsection{Results}
In the following we examine the behaviour of three models, 
with two different initial soundspeeds $c_{s,\infty}$ and
the same asymptotic velocity $v_\infty$.  
The prefix ${\tt sp}$ is used to denote "${\tt s}$trongly
${\tt p}$erturbed" which means that 
the initial asymptotic pressure is lowered by two orders
of magnitude with respect to the equilibrium
value. This is done with the main
  purpose of producing models with even lower
  temperatures.\\
\paragraph*{Measured physical quantities}
In addition to the primitive variables
provided by the code,  we calculate several physical quantities: 
the accretion rate in Eddington
units, $\dot{M}$, the luminosity in Eddington units, $L$,  
the radiation equivalent temperature, $T_{\rm rad}=\left(E_{\rm r}/a_{\rm rad}\right)^{1/4}$,
the fluid temperature, $T_{\rm fluid}$, the effective adiabatic index $\gamma_{\rm eff}$:
\begin{center}
\begin{figure*}
{
\includegraphics[width=8.0cm,angle=0]{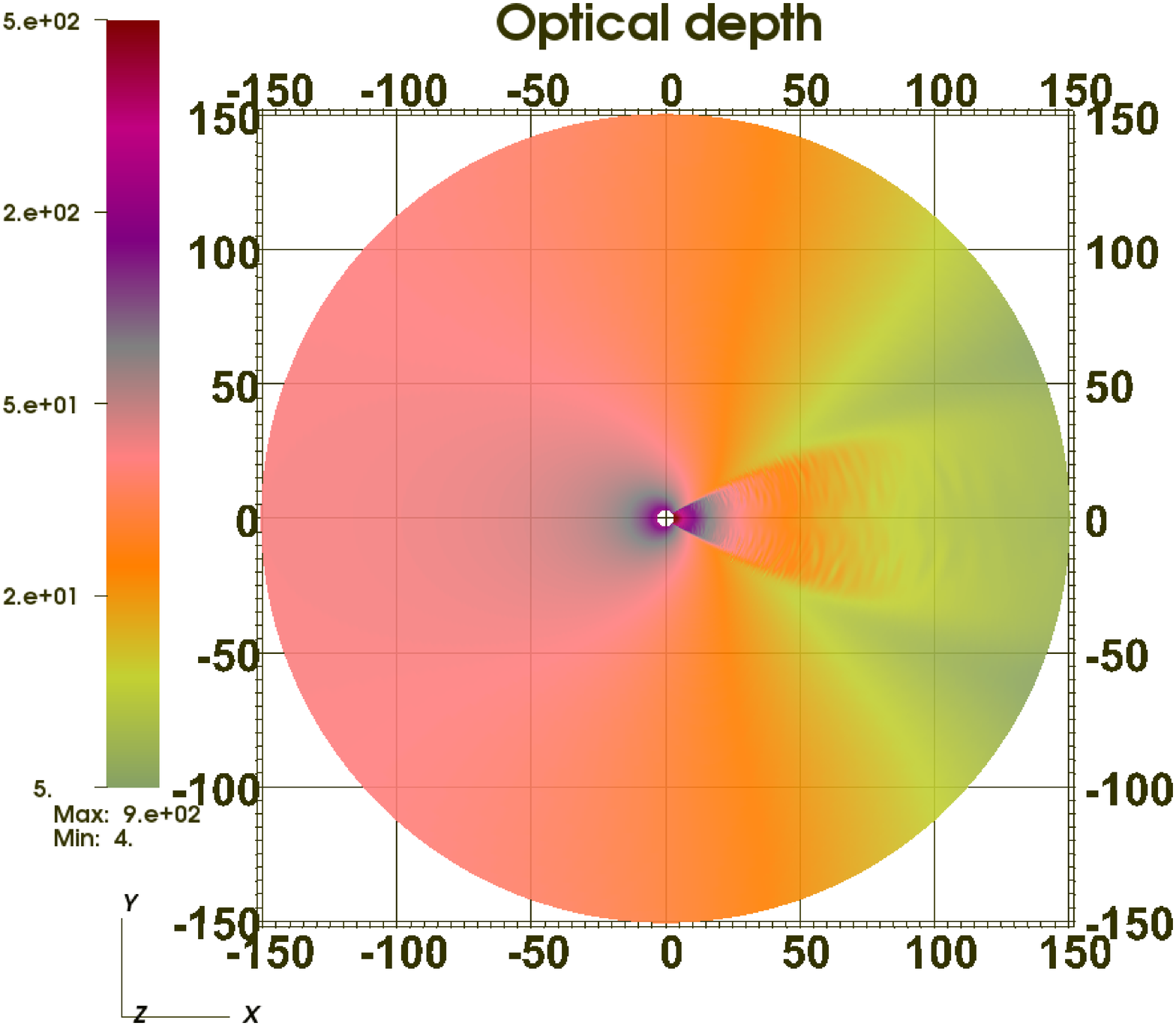}
\hskip 0.5cm
{\includegraphics[width=8.0cm,angle=0]{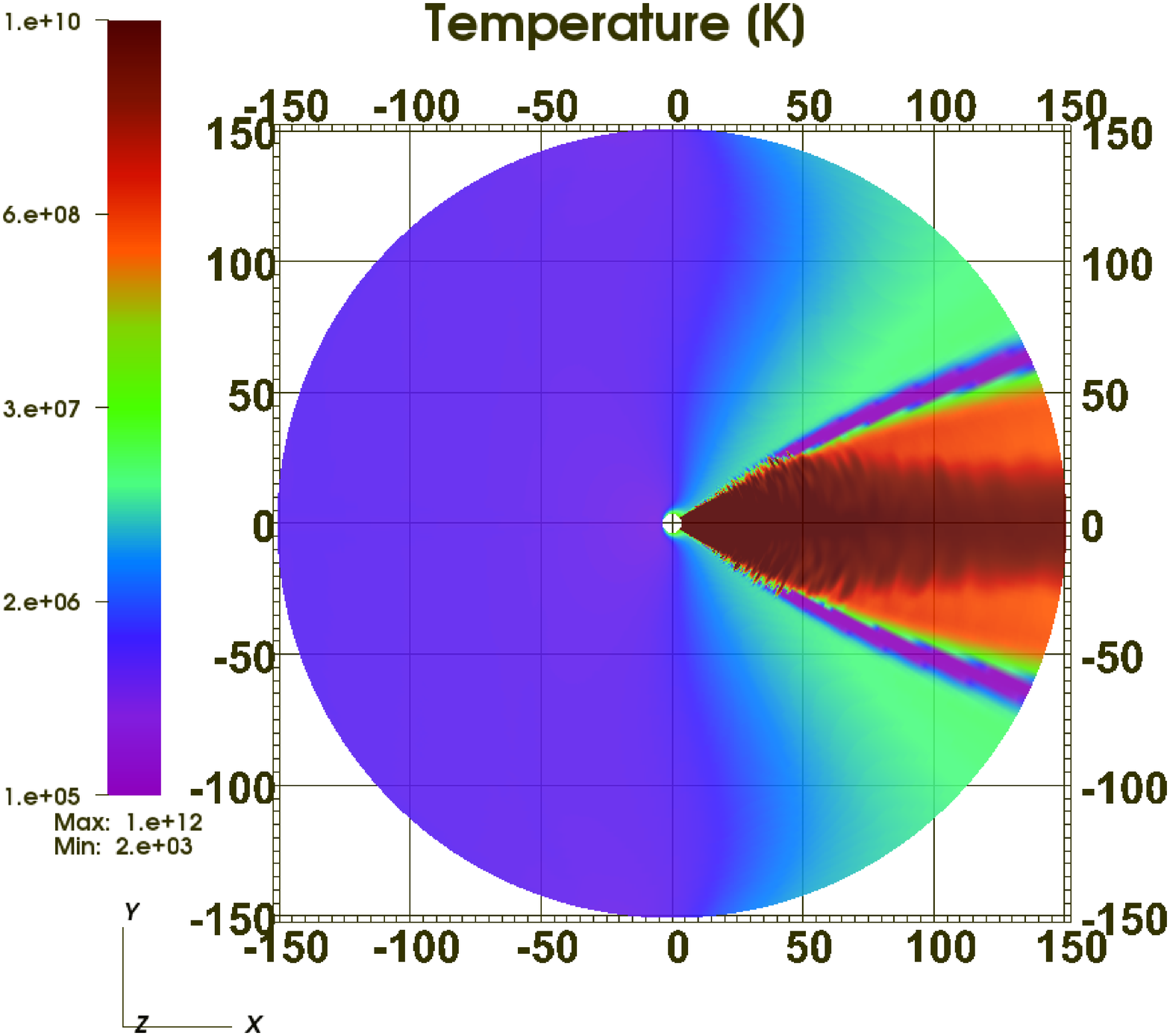}}
\includegraphics[angle=0,width=8.0cm]{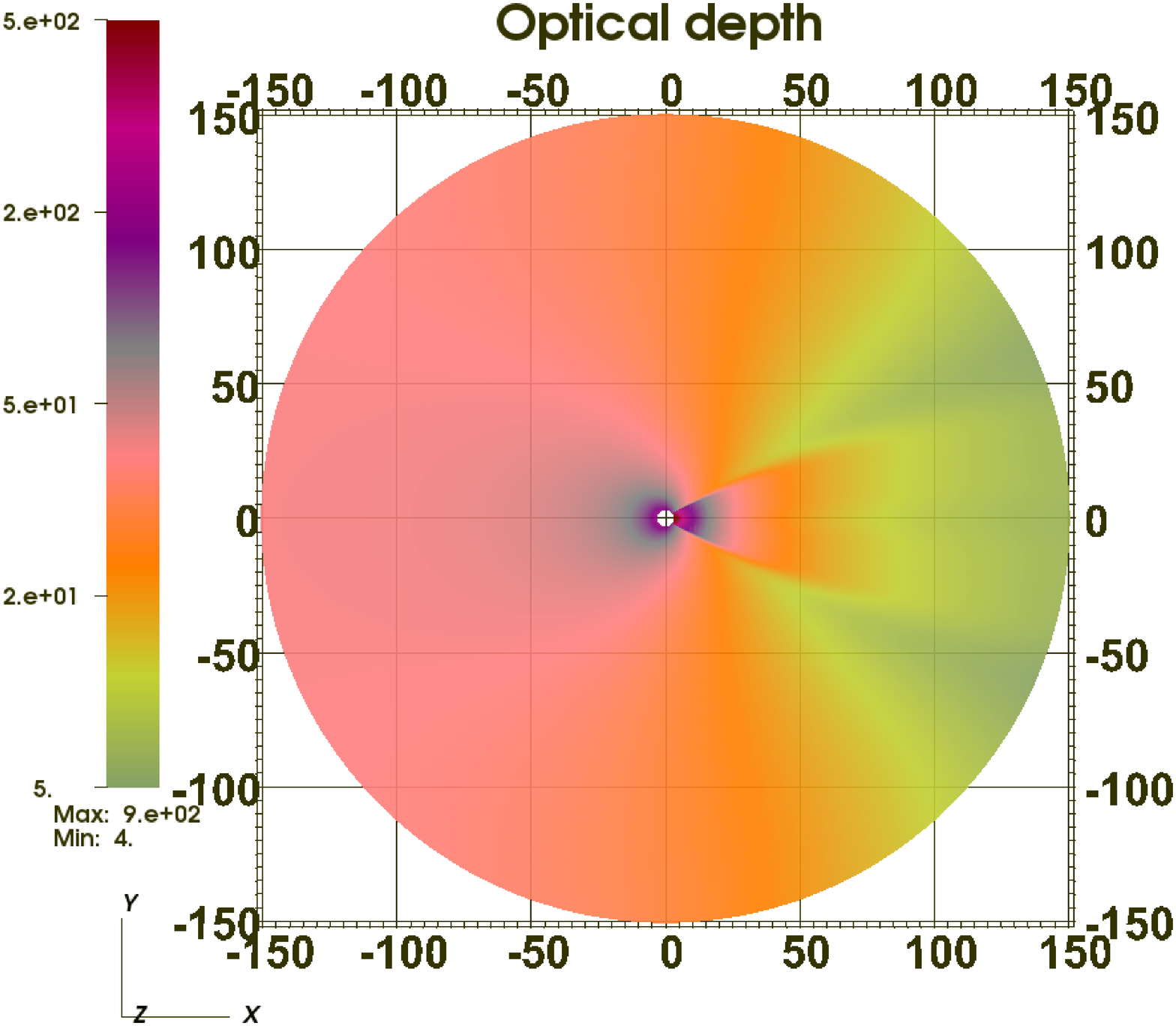}
\hskip 0.5cm
{\includegraphics[angle=0,width=8.0cm]{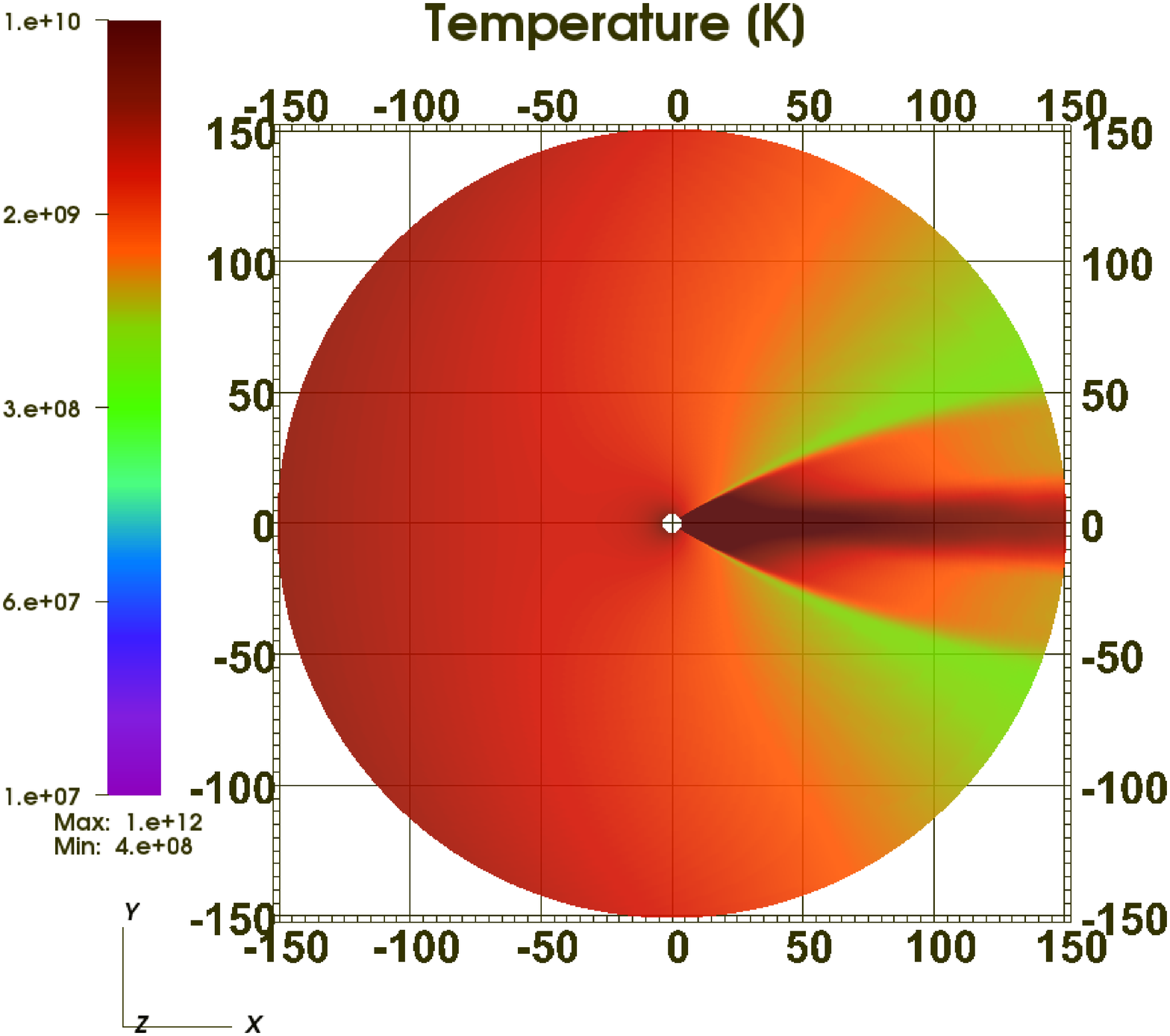}}
\caption{{\sc 2D Optical depth and fluid temperature of
    perturbed model ${\tt sp.V07.cs03}$ 
  and unperturbed  models ${\tt V07.cs03}$-}} 
Both models are shown at stationary state: (Top)
$t=7.71\times 10^4\,M$ for model  ${\tt sp.V07.cs03}$  and (Bottom)
$t=5.98\times10^4\,M$ for model ${\tt V07.cs03}$.
\label{fig:2DTprsp}
}
\end{figure*} 
\end{center}

\be
\gamma_{\rm{eff}}=\frac{5/2+20q+16q^2}{(3/2+12q)(1+q)} \,\,\,\text{with   }\,\,q={\cal P}_{\rm r}/P\,,
\ee
 and the local Mach number $\mathcal{M}$:
\be
\mathcal{M}=\frac{\Gamma\, \sqrt{v_i v^i}}{\Gamma_{c_s}
  c_s}\,\,\,\text{with   }\,\, c_s=\sqrt{\frac{\gamma
    P}{h \rho }} = \sqrt{\frac{\gamma
    P}{\rho +\frac{\gamma}{\gamma-1} P}}\,.
\ee
Moreover, as discussed in paper{\sc I}, we define an effective BHL
luminosity efficiency $\eta_{_{\cal BHL}}$, that takes
into account the injected energy at infinity as 
\be
\eta_{_{\cal BHL}}=\frac{L} {\dot{M}_{\rm acc} c^2 +
\frac{1}{2}\dot{M}_{\infty} v^2_{\infty}}\,.
\label{eq:defeta}
\ee

\begin{table*}
\caption{\label{tab:newruns} Representative quantities of
  the considered models after quasi-stationary state has
  been reached. 
The columns report the model name,
the average radiation temperature,
the average effective adiabatic index,
the accretion rate, 
the luminosity and 
the radiative efficiency, all of them computed after a
 quasi-stationarity state had been reached.
See text for definition of these quantities.}
%
\begin{tabular}{lccccc}
\hline \hline
Name & $\langle T_{\rm rad}\rangle$[K] & $\langle\gamma_{\rm eff}\rangle$& 
$\dot{M}/\dot{M}_{\rm Edd}$ &$L/L_{\rm Edd}$&$\eta_{_{\cal BH}}$\\
\hline
${\tt V07.cs03}$   & $5.6\times 10^5$  &1.333 & 132 & 0.939  &$6.9\times 10^{-3}$\\
${\tt sp.V07.cs03}$ & $5.6\times 10^5$ &1.334 & 135 & 0.943  &$6.8\times 10^{-3}$\\
${\tt sp.V07.cs05}$ & $4.3\times 10^5$ &1.333 & 62  & 0.484  &$9.0\times 10^{-3}$\\
\hline
\end{tabular}
\end{table*}

\begin{figure}
{\includegraphics[clip=true,width=8.0cm]{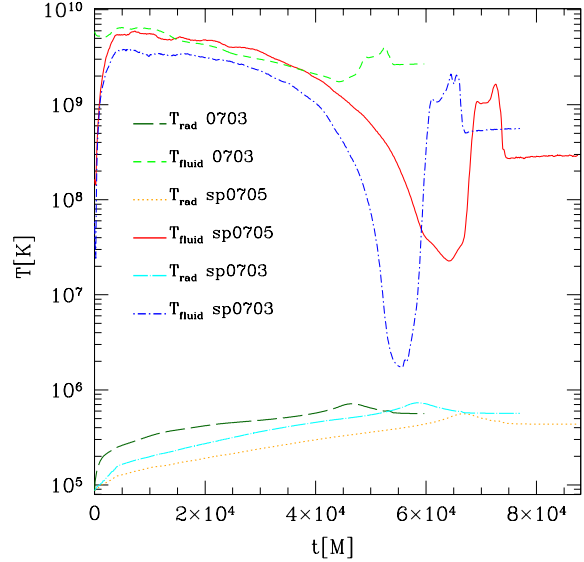}}
\caption{\label{fig:Ten}{\sc Comparison of fluid and radiation temperatures} - Volume weighted averages according to Eq.~\eqref{eq:ave}
for all three BHL models \textcolor{red}{${\tt sp.V07.cs05}$}, 
\textcolor{blue}{${\tt sp.V07.cs03}$} and
\textcolor{green}{${\tt V07.cs03}$} as a function of
time. Temperatures are given in Kelvin. 
}
\end{figure}

\begin{center}
\begin{figure*}
{
\includegraphics[angle=0,width=8.0cm]{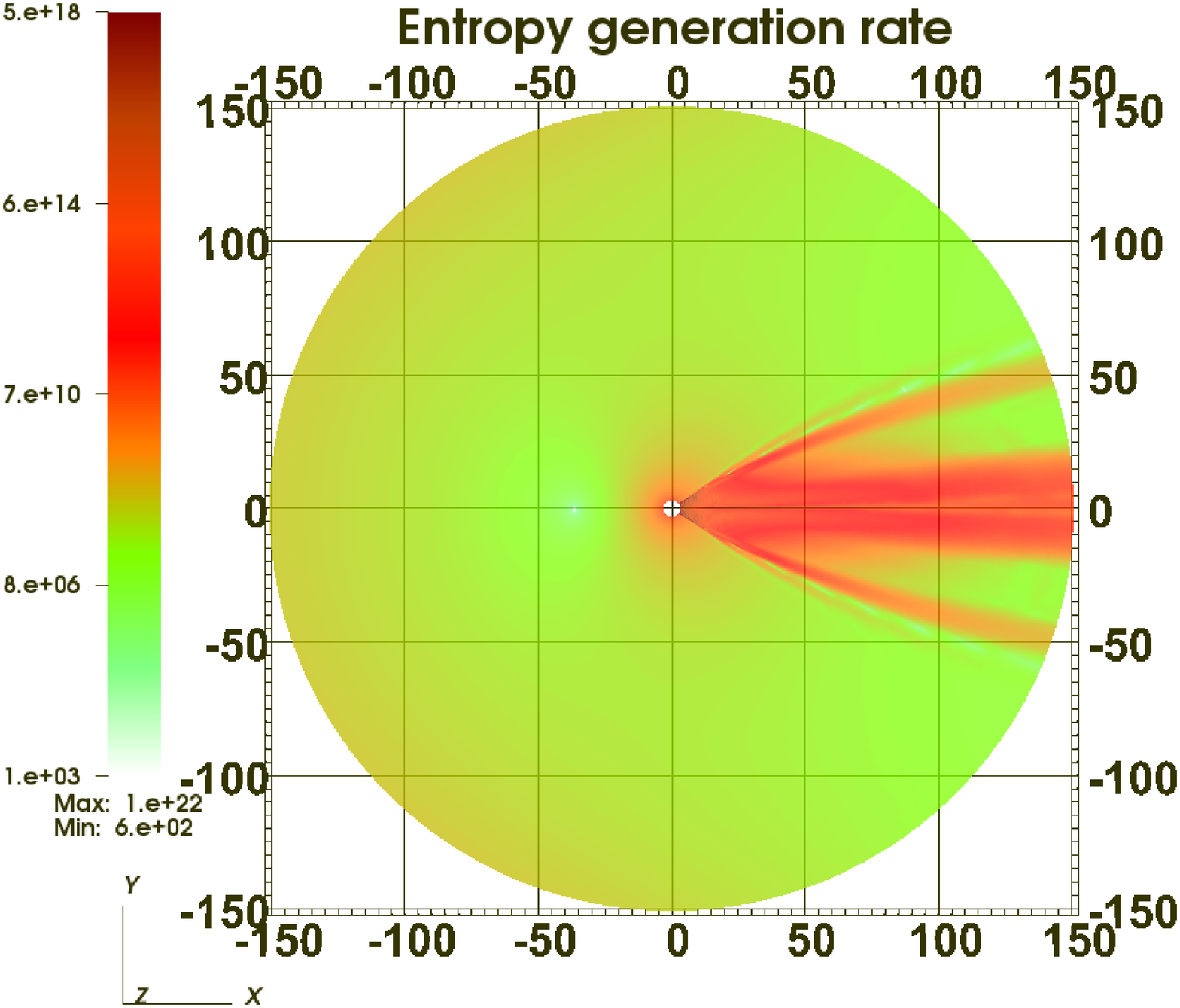}
\hskip 0.5cm
\includegraphics[angle=0,width=8.0cm]{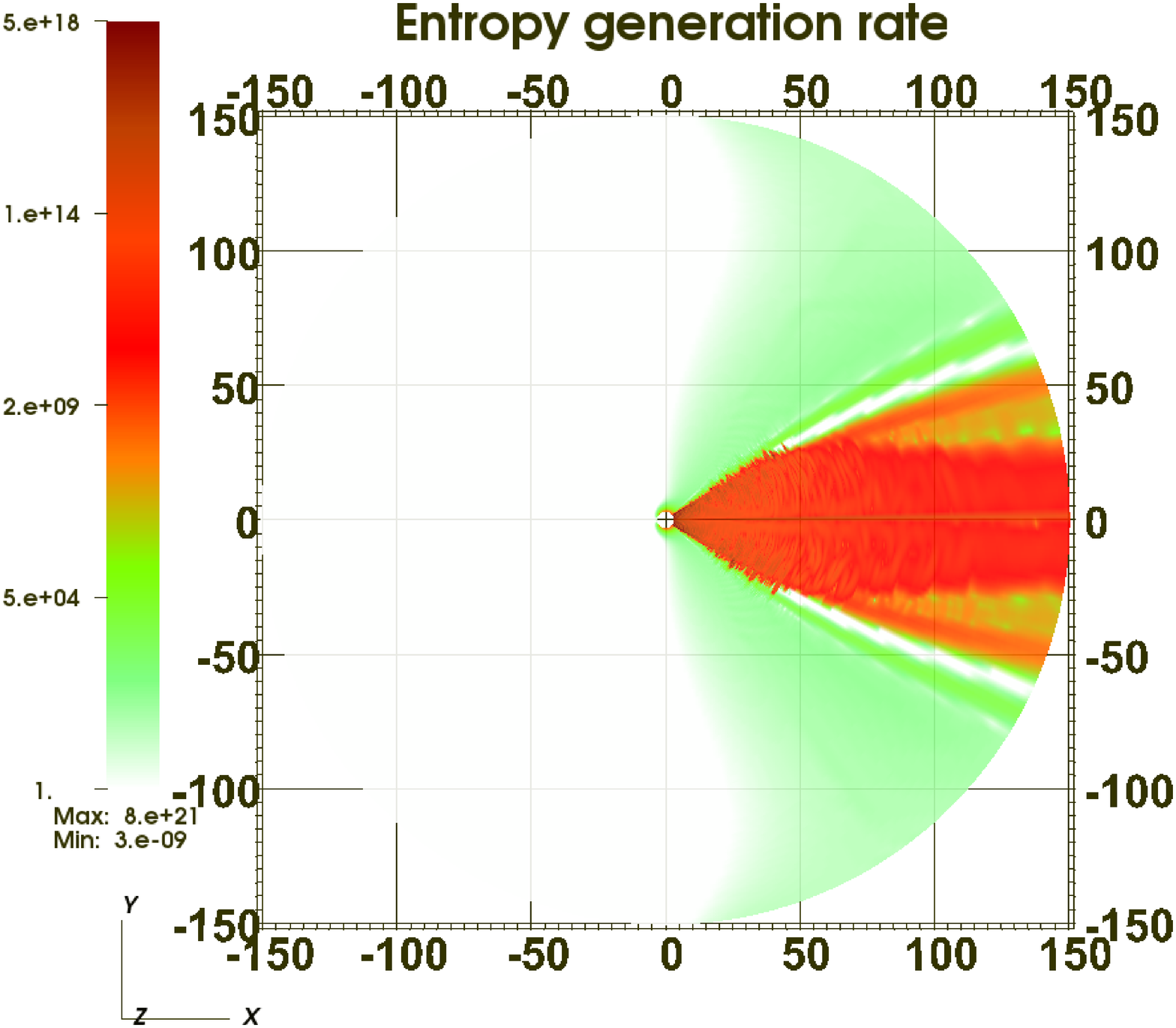}
\caption{\label{fig:2DEntsp} {\sc 2D Entropy generation rates for ${\tt V07.cs03}$ and  ${\tt sp.V07.cs03}$ } -}
Distribution map of $\nabla_\nu S^\nu$  for
model ${\tt V07.cs03}$ (left panel) 
at time $t=5.98\times10^4\,M$ and for
model  ${\tt sp.V07.cs03}$ (right panel) at time $t=7.71\times 10^4\,M$.}
\end{figure*}
\end{center}

We measure several quantities $Q$ as volume weighted averages over all grid 
elements $i$, thus defining the pointy brackets as  
\be
\langle Q \rangle = \frac{1}{\sum^N_i r_i\text{d}r_i\text{d} \phi_i } \left(\sum^N_i Q_i \, r_i\text{d}r_i\text{d} \phi_i\right)\,.
\label{eq:ave}
\ee
The rate of entropy generation is measured according to Eq.~\eqref{eq:entrcons_npf3}, which is an appropriate approximation for a coupled photon 
fluid plasma in a quasi-stationarity state.
When we extract the luminosity, we choose a surface of constant optical depth $\tau \gsim 10$.
We argue this is reasonable because only if $\tau \gg 1$  the system is still
in a regime where the approximation with the diffusion limit is valid.
This is also realistic, when thinking of actual observations, 
where measurements are taken at constant\footnote{In the
  case of stars, for instance, this is usually taken as
  $\tau \sim 2/3$.} $\tau$.
For a discussion of how the luminosity estimates change with respect to paper{\sc I}, see Appendix~\ref{appendixA}.
Suffice it to say that this luminosity is a tracer of the outwards radial fluxes and that
different possible extractions agree within the current error bars.\\
The optical depth $\tau$ is computed in post processing as in paper{\sc I}
\be
\label{eq:tau}
\tau=\int_0^L (\chi^t + \chi^s) ds \,,
\ee
where we have assumed a constant characteristic lengthscale $L=10$.
All other quantities are standard
and reported in ${\rm cgs}$ units.
Selected results are shown in
Figs.~\ref{fig:2DTprsp} to~\ref{fig:2DEntsp} and
summarised in Tab.~\ref{tab:newruns}.\\ 

\paragraph*{BHL dynamics dominated by radiation quantities}
First of all, we note that the qualitative dynamics of
all BHL models 
is the same as 
described in paper{\sc I} and 
can be summarised as follows [see also
  \citet{Petrich89,Font98a,Donmez2010} for the hydrodynamics case
  and \citet{Penner2011} for the magnetohydrodynamics one].
Initially, a narrow, hot shock cone
forms downstream of the accretor and the plasma is
fluid-pressure dominated. 
Progressively, the radiation field builds up strength
until the radiation pressure becomes similar to the fluid
pressure. At this point, the shock cone becomes unstable, 
oscillating from one side of the accretor to the other,
until it finally reverses into the upstream domain as a bow shock.
From now on the radiation pressure exceeds the fluid
pressure, the effective adiabatic index approaches the
value $\sim 4/3$
 and, at the same time, the  density (and
correspondingly the optical depth) decreases in most
parts of the numerical domain. 
After the upstream shock has moved out of the numerical
domain and expelled  a significant amount of mass,  
a new, low density equilibrium is formed in which there
is a smaller shock cone in the downstream region 
(this is illustrated well by Fig.~3 of paper{\sc I},
which shows 
a comparison of BHL flows with and without the radiation
field).\\ 
The central improvement over paper{\sc I} is shown in
Fig.~\ref{fig:2DTprsp}, 
where the two-dimensional maps of 
the optical depth (left) and of
the fluid temperature (right)
are shown for two different
models, both of them at the final quasi-stationary state. 
The top panels, in particular,  show that for
the strongly perturbed model \textcolor{blue}{${\tt sp.V07.cs03}$}
large parts of the upstream region settle down to temperatures of the order
$T_{\rm fluid}\lsim 10^6$\,K, a value which could not
  be reached before due to the stiffness of the equations.
The corresponding unperturbed model, \textcolor{green}{${\tt V07.cs03}$},
shown in the bottom panels,
has upstream temperatures as high as $T_{\rm fluid}\lsim5\times
10^9$\,K, while both of the models have significantly high
optical thickness.
It is important to note that at this stage of the
evolution, namely after the "reversal" of the shock cone,
the effective adiabatic index of all three models is very 
close to $\gamma_{\rm eff}\sim 4/3$,
and therefore behaving like an effective
photon fluid.\\
Additional understanding of the thermodynamics of the
models is achieved
if we look at the time evolution of
the averaged fluid and averaged radiation temperatures, which are
plotted in Fig.~\ref{fig:Ten}.
There are three points worth noting.
First of all, for each model, the two temperatures
$T_{\rm rad}$ and $T_{\rm fluid}$ 
differ by many orders of magnitude, suggesting that, at
least globally, 
there is a strong deviation from thermal
equilibrium within the fluid. 
Secondly, the fluid temperatures 
of the models \textcolor{blue}{${\tt sp.V07.cs03}$} and \textcolor{green}{${\tt V07.cs03}$} 
are also significantly different, in spite of the dynamics being
very similar
(this is discussed in the next \S). 
Finally, $T_{\rm rad}$ shows a smooth evolution, whereas
$T_{\rm fluid}$ exhibits a strong dip, reaches a minimum, and
heats up again afterwards. 
When the large size and hot ($T_{\rm fluid}\gg 10^{10}$\,K) shock cone
reverses, the density downstream of the accretor
becomes small, yet the pressure remains high. 
A smaller size high temperature shock cone forms in the downstream region, 
as visible
in  the right panels of Fig.~\ref{fig:2DTprsp},
with $T_{\rm fluid}$-average
being dominated by the high values within the shock cone.
Thus, the fluid behaves like an effective photon fluid 
of temperature $T_{\rm rad}\lsim 10^{6}$\,K, but in the shock cone
no thermalization is possible and
the fluid temperature vastly exceeds the radiation temperature.
In order to corroborate this description,
we measure the entropy generation rate $\nabla_\nu S^\nu$ as an effective
tracer of dissipative processes
(see Appendix \ref{appendixA}).  The entropy generation rate (
Fig.~\ref{fig:2DEntsp} for the two models ${\tt
  V07.cs03}$ and  ${\tt sp.V07.cs03}$)
is maximum in the shocked region downstream of the
accretor, where the fluid is very 
far from thermalization, even though the
dynamics is otherwise stationary.
Only in the upstream regions where $\nabla_\nu
S^\nu$ is very small (cf. the
white region of Fig.~\ref{fig:2DEntsp}), are the two temperatures
$T_{\rm rad}$ and $T_{\rm fluid}$ similar.

\begin{center}
\begin{figure*}
{
\includegraphics[clip=true,width=8.0cm]{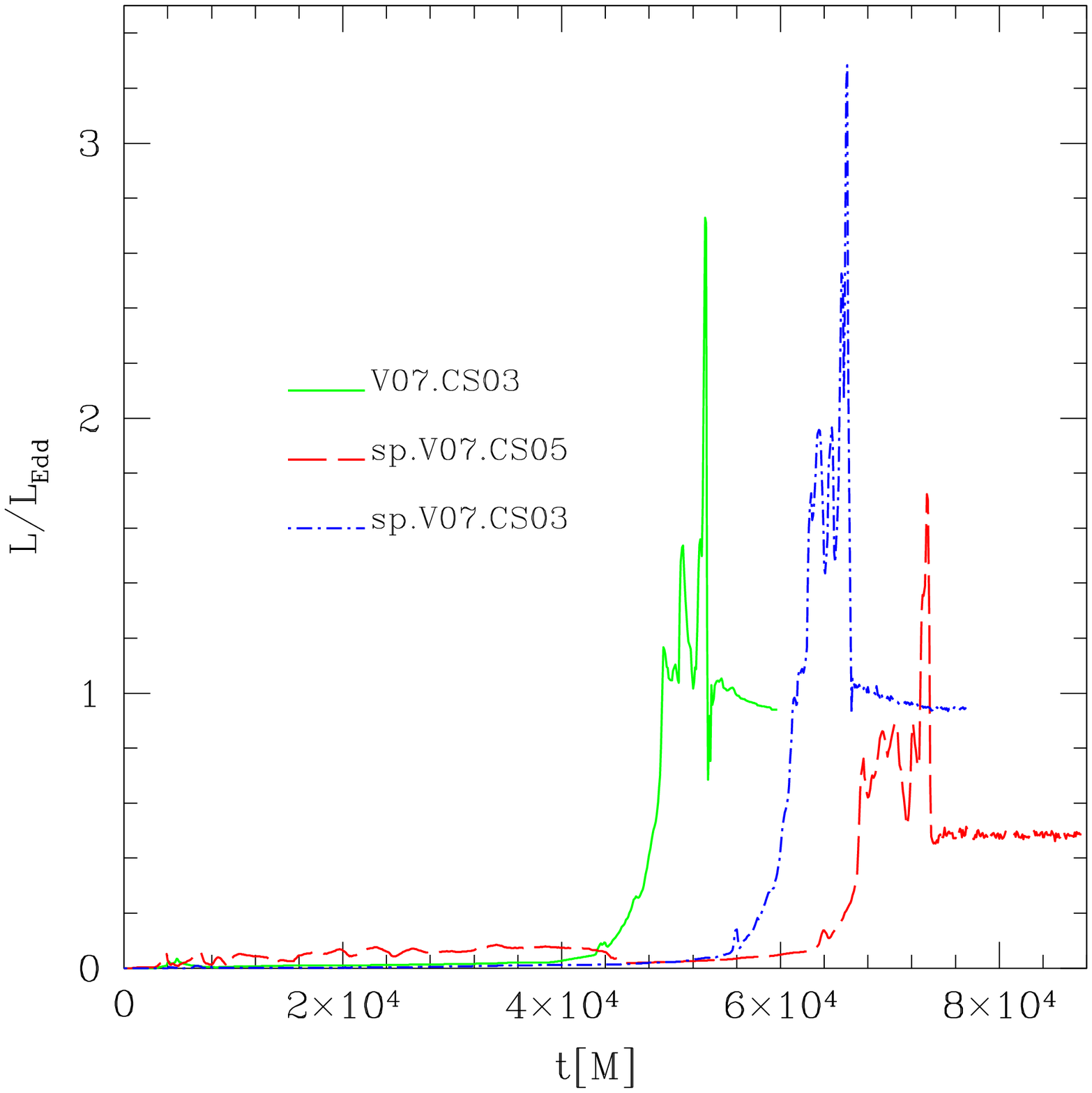}
\hskip 0.5cm
{\includegraphics[clip=true,width=8.0cm]{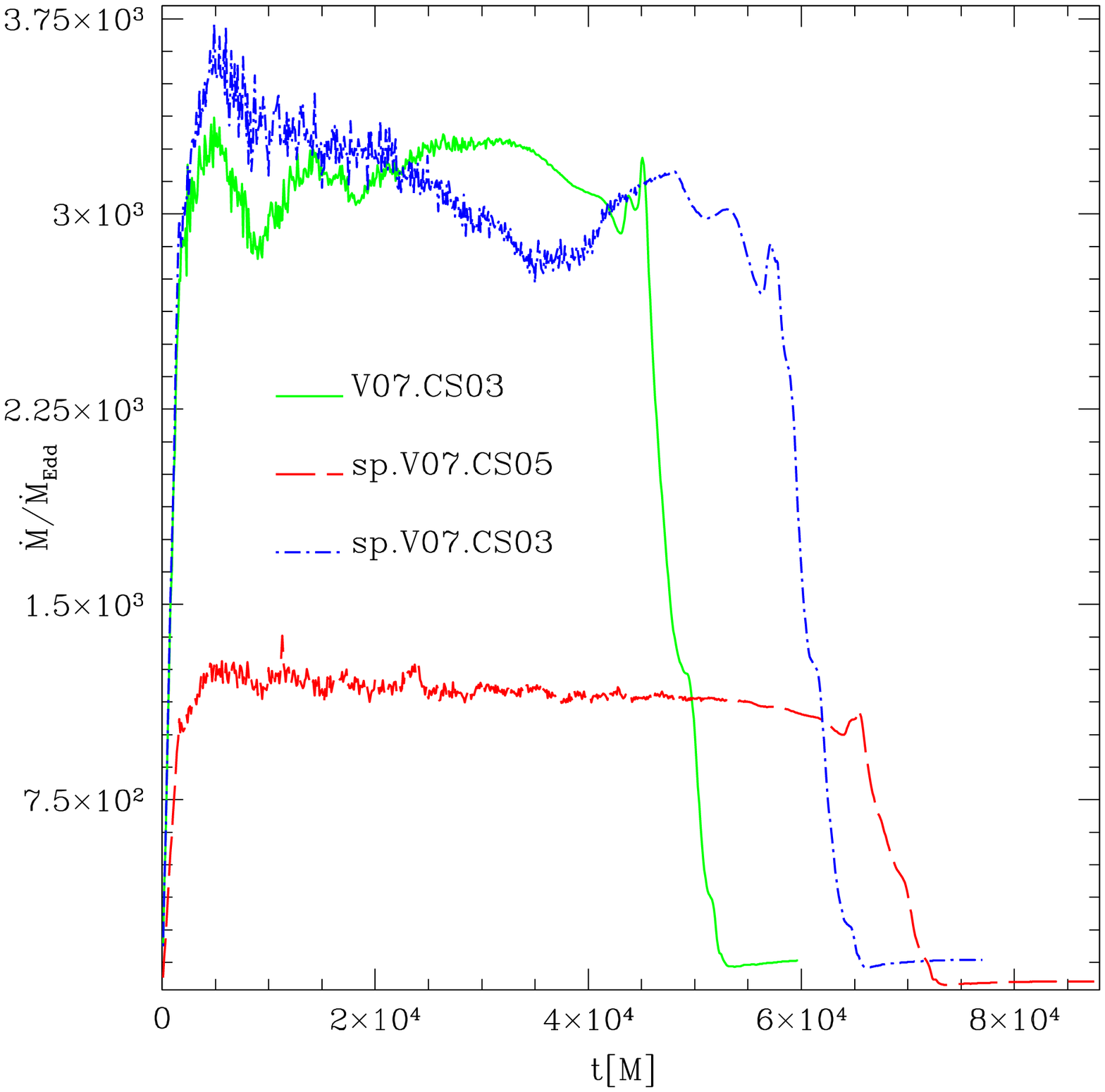}}
\caption{{\sc Time evolution of perturbed BHL models \textcolor{red}{${\tt sp.V07.cs05}$}, \textcolor{blue}{${\tt sp.V07.cs03}$} 
and \textcolor{green}{${\tt V07.cs03}$}}-}  
(Left panel) luminosity $L$ extracted at constant optical depth $\tau\ge10$,
(Right panel) accretion rate $\dot{M}$ as a function of time in Eddington units. \label{fig:spbhcompareL}
}
\end{figure*}
\end{center}
To further illustrate the effects of radiation induced dynamics,
we measure two crucial 
parameters of accretion flows, namely the luminosity and
the accretion rate, both of them reported in Eddington units.
The computation is performed by directly integrating  the escaping
radiation fluxes $f^r$ (Eq.~\ref{eq:luminosity}),
and the infalling mass fluxes at the innermost grid-point, respectively.
We plot the time evolution of the
luminosity and of the accretion rate
on the left and on the right panels of Fig.~\ref{fig:spbhcompareL}
for all three models. 
The important aspects of this figure are that
(i) there is a transient peak in the luminosity evolution, corresponding
to the point in the dynamics where the shock cone is
momentarily dissipated away; 
(ii) the final luminosity is sub-Eddington for all cases;
(iii) 
the luminosities of the models
\textcolor{blue}{${\tt sp.V07.cs03}$} and
\textcolor{green}{${\tt V07.cs03}$} converge towards the
same value; and 
(iv) the higher sound speed (correspondingly the lower
asymptotic Mach number $\mathcal{M}_\infty$)  of
\textcolor{red}{${\tt sp.V07.cs05}$} leads to smaller
luminosity. On the other hand, the corresponding
  accretion rates are substantially super-Eddington, with
  final values of $\dot{M}/\dot{M}_{\rm Edd}$
in the range $[62,135]$,
and confirming the advection dominated nature
  of BHL accretion flows. The relaxed luminosity efficiency
$\eta_{_{\cal BHL}}$ of the models together with all radiation quantities are listed in Tab.~\ref{tab:newruns}.

\begin{figure}
\vspace{-3.cm}
{\includegraphics[angle=0,width=8.0cm]{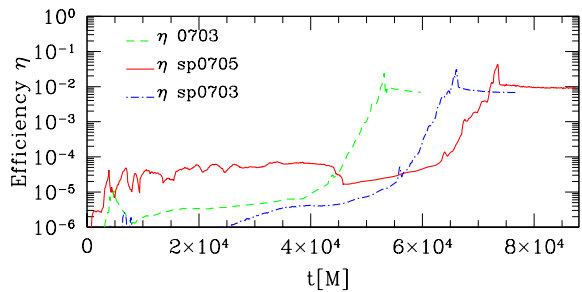}}
\caption{\label{fig:meta}{\sc Radiative efficiency} -
  Comparison of the radiative efficiency $\eta_{_{\cal BHL}}$
as a function of time. }
\end{figure}

\paragraph*{The role of fluid temperature}
It is interesting to note that the strongly perturbed
model \textcolor{blue}{${\tt sp.V07.cs03}$} converges towards a
final state that is very similar to that of its
unperturbed counterpart \textcolor{green}{${\tt V07.cs03}$}. 
This is observed in the accretion rate, $\dot{M}$,
plotted on the right panel of Fig.~\ref{fig:spbhcompareL}, in
the luminosity, in the radiative efficiency, $\eta$
(cf. Fig.~\ref{fig:meta}), in the optical depth, $\tau$
(cf. Fig.~\ref{fig:2DTprsp}), and in the   
 radiation temperature, $T_{\rm rad}$
 (cf. Fig.~\ref{fig:Ten}) of these two models. 

This effect remained obscured in paper{\sc I}, 
due to the fact that the previous 
criterion for the luminosity extraction was spuriously affected by
boundary effects\footnote{See
discussion in Appendix \ref{appendixA}.}.

While the radiation quantities converge for the two
models \textcolor{blue}{${\tt
    sp.V07.cs03}$}  and 
\textcolor{green}{${\tt V07.cs03}$}, 
the quantities more directly
related to the fluid properties do not.
For example, the
fluid temperature, the Mach number and the entropy generation are
neither qualitatively and certainly not quantitatively
the same. 
In addition, the radiation dominated regime is reached at earlier times
 for \textcolor{green}{${\tt V07.cs03}$} as it has higher
 $T_{\rm fluid}$ and thus higher 
thermal conductivity (cf. Appendix~\ref{appendixA}).

From the consideration above, it stands to reason that the radiation
temperature and 
the matter density (conversely, the optical depth) are the quantities
affecting the  dynamics most. 
This means that bremsstrahlung cannot  not be  
a dominant process, since it is a temperature dependent
radiation interaction. 
This is confirmed by the fact that,
when looking at the respective opacities, Thomson
scattering dominates over 
bremsstrahlung by several orders of magnitude. 
We also note that in some portions  
of the grid, the discrepancy between 
$T_{\rm fluid}$ and $T_{\rm rad}$ is very large, implying
that the assumption of LTE is not valid there
 (cf. the red region of Fig.~\ref{fig:2DEntsp}). 
This is consistent with the fact
that full thermalization in general is very
hard to accomplish in dynamical environments of moderate
density.

\paragraph*{Further comments}
We had already pointed out in paper{\sc I} that models
with initial high Mach number, $\mathcal{M}_\infty$, are
characterized by a luminosity that is dominated by the
emission at the shock front, rather than
by accretion-powered luminosity. This is also confirmed
by the relative comparison between 
\textcolor{blue}{${\tt sp.V07.cs03}$} and
\textcolor{red}{${\tt sp.V07.cs05}$}, the former
having a larger Mach number and a higher luminosity (left
panel of Fig.~\ref{fig:spbhcompareL}).

Finally, we would like to comment about what has been dubbed the
"`flip-flop"' instability in BHL accretion flows,
and whose physical nature is still a matter of
debate~\citep{Foglizzo2005}. 
While we do not see this instability in our models
(neither in paper{\sc I} nor in the present), we have observed
that  during the "`shock-reversal"' 
strong, although transient, 
oscillations in the shock cone can appear.  
However, we suspect that this effect can be partly
attributed to the numerics, since 
the use of the IMEX in combination with a
higher order Runge Kutta (order $3$ instead of $2$), 
alters the behaviour of this oscillation slightly. 
An extended analysis through three dimensional simulations
would be needed to establish the potential relation 
of this  oscillatory behaviour with the eventual
development of the flip-flop instability.

\section{Concluding remarks}
\label{sec:conclusions}

In this paper, we have revisited the optically thick,
thermal radiation transfer in GR. First, we  addressed
the numerical problem of stiff source terms; proposed a
numerical treatment, 
implemented and verified it. As we chose an IMEX
Runge-Kutta scheme, we needed to isolate 
the principal stiff parameters, 
which were found to be the (density-weighted) opacities.
After applying the new IMEX method to the one-dimensional
problem of spherical accretion, we compared our results
with those obtained earlier by \citet{Nobili1991} and
found good agreement.  
In this spherical, stationary scenario the current
formulation of the GR-RHD equations is fully applicable
as long as the solution remains optically thick.
We remark that there is not a
unique stiffness threshold, valid for any physical
scenario, at which the purely explicit scheme
fails and the IMEX becomes necessary. 
In the case of a purely explicit RK scheme, when the
source terms become stiff, it is possible to a certain extent to lower the
$\cfl$ factor and obtain a stable evolution. However, the
stiffness parameter can become very large, so the time-step very
tiny. 
That is of course inefficient, and resorting to 
a stiff solver is the only way out.
In general, if a problem can be
solved with a purely explicit RK scheme, this is to be
preferred as it is CPU-faster. However, we believe that
most non-trivial radiation applications will exhibit
stiffness and lead to code crashes with standard explicit
RK schemes. 
 
We then revisited the Bondi Hoyle Lyttleton accretion in
2D for an astrophysical, dynamical problem. 
Here, we could show that: 
\begin{itemize}
\item
The IMEX scheme allows us to evolve models with
realistic choice of parameters, of the order of
$T\sim10^6$\,K; 
\item the dynamics of the
flow are significantly  affected
by the radiation pressure, yielding 
super-Eddington accretion rates in the range
$\dot{M}\sim[62,135]\dot{M}_{\rm Edd}$
and Eddington limited luminosities;
\item
the fluid and the
radiation depart strongly from 
thermal equilibrium in shocked regions, particularly in
the shock cone downstream of the accretor. 
\end{itemize}
Our analysis has substantially benefited from the ability of
our scheme 
to treat stiff source terms. However, we should 
also state a few words of caution as to the current
shortcomings and necessary future improvements of our
scheme:
\begin{itemize}

\item The optically thin regime cannot be treated yet, and 
  further steps are required to incorporate 
the variable-Eddington factor approach.
 \item Temperatures of order $T<10^5$\,K, as they appear
   in small regions of the domain, 
       require the inclusion of bound-free opacities,
       which are currently neglected.
 \item  The only dissipative mechanism is currently
   thermal conductivity.  Other types of viscosity such
   as an effective viscosity related to 
   magnetic turbulence would be beneficial. Coupling the
   current equations to MHD represents another direction
   of future research.
\item 
Since we currently cannot extract the luminosity 
in regions where the optical depth is low, 
we must trace a geometrical surface of constant
$\tau\geq1$. However, it remains an uncertainty as to
where such a surface should be placed, and the computed
luminosities are therefore affected by at least one order
of magnitude uncertainty.

\end{itemize}
Even in the presence of these limitations, our analysis
may become relevant 
for the study of merging
supermassive black-hole binaries, which have been
attracting a lot of interest for the possible joint
measure of electromagnetic and gravitational wave signal
(in the context of multi-messenger astronomy).
Neglecting the back-reaction of
radiation onto matter, \citet{Farris:2009mt} already
considered the BHL solution in a binary system, finding
that luminosities as high as $10^{43}{\rm erg}\,s^{-1}$
can be obtained in a hot gas cloud 
of
 temperatures $T\sim 10^6$\,K. 
Such estimates are compatible with our calculations, but
a dedicated work will be presented in the future.

\section{Acknowledgements}
CR wishes to thank Nico Budewitz for his helpfulness in HPC support and Aaryn Tonita for his help with the HDF5 I/O.
DA thanks E. Schnetter for
his guidance during the implementation of the RKIMEX method in the MoL thorn.
We are grateful to the anonymous referee whose comments helped improve
the clarity of the manuscript.
We express our gratitude to Luca Zampieri for providing
us with the data shown in Fig.~\ref{fig:test_michel}.
The simulations were performed at the {\it datura}
cluster of the AEI and at SuperMUC at the LRZ M\"unchen. 
This work was funded in part by the SFB Transregio 7 of
the DFG. CR acknowledges funding by the "International Max Planck Research School". OZ
acknowledges funding by the 
European Union's Seventh Framework Programme
(FP7/2007-2013) under the research  
project \textit{STiMulUs}, ERC Grant agreement no. 278267. 

\appendix
\section[]{Entropy generation rate and luminosity computation}
\label{appendixA}

\begin{figure}
{\includegraphics[clip=true,width=8.0cm]{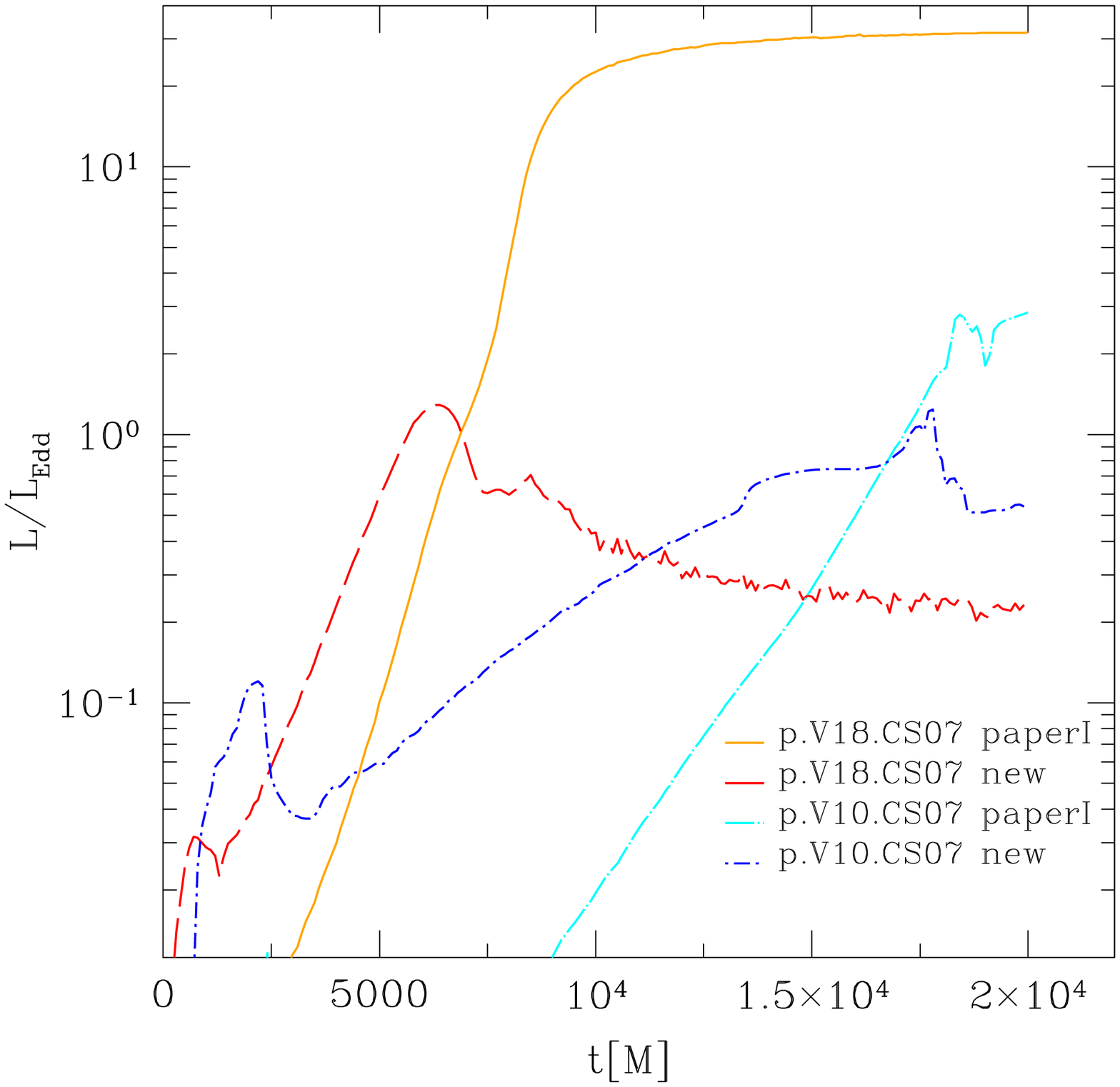}}
\caption{{\sc Comparison of luminosity extraction of perturbed BHL} - \textcolor{red}{${\tt p.V18.cs07}$} and
\textcolor{blue}{${\tt p.V10.cs07}$}. Extracting the luminosity at $\tau\ge10$ leads to different light curves, these curves
are labelled "new". In paper{\sc I} we had used the criterion $\tau\ge1$.
\label{fig:Lcomp}
}
\end{figure}

In the framework of Eckart's formulation of Relativistic
Standard Irreversible Thermodynamics~\citep{Eckart40c}, the entropy
current is given by
\begin{equation}
\label{entropy-current}
\mathcal{S}^\mu=s \rho u^\mu + \frac{q^\mu}{T}\,,
\end{equation}
where $q^\mu$ is the heat flux, 
$s$ is the entropy per unit mass, and $T$ is the
temperature of the fluid. 
The heat flux is given by the relativistic form of Fourier law, namely
\citep{Israel76}
\be
\label{heat-flow}
q_\mu=-\lambda T(h_\mu^\nu \nabla_\nu\ln T + a_\mu)\,,
\ee
where $a^\mu$ is the four-acceleration of the fluid, 
$\lambda$ is the thermal conductivity and
$h_{\mu\nu}=g_{\mu\nu}+u_\mu u_\nu$ is the projector
operator in the space orthogonal to the four-velocity $u^\mu$.
Under the assumption that the 
fluid has vanishing shear and vanishing bulk
viscosity, 
the entropy-generation rate that follows from
\eqref{entropy-current} and \eqref{heat-flow}
is given by 
\be
\label{eq:entrcons_npf}
T\nabla_\mu \mathcal{S}^\mu=\frac{q^\mu q_\mu}{\lambda T}\,.
\ee
We recall that the thermal
conductivity is related to the opacity.
For instance, the thermal
conductivity computed using the ordinary diffusion approximation of
stellar interiors is given by $\lambda=(4/3) a_{\rm rad}c
T^3/\chi^s$~\citep{Schwartz1967}. 
Under the assumption that the matter plus
radiation fluid behaves as a single fluid with effective
pressure and energy density given by $P_{\rm eff}=P+{\cal
P}_{\rm r}$, $e_{\rm eff}=e+E_{\rm r}$, 
the four acceleration $a_\mu$ can be computed from the Euler
equations as
\be
a_\mu=-\frac{h^\nu_\mu\nabla_\nu P_{\rm eff}}{e_{\rm eff}+P_{\rm eff}}\,.
\ee
When quasi stationary configurations are reached, the
terms containing time derivatives can be neglected with
respect to those  containing spatial derivatives, and 
after replacing $q^\mu$ into
Eq.~\eqref{eq:entrcons_npf} we obtain
\bea
\label{eq:entrcons_npf3}
\nabla_\mu
\mathcal{S}^\mu&&\approx\frac{\lambda}{T^2}\bigg[(g^{rr}+\Gamma^2 (v^r)^2)
  (\partial_r T)^2+ \nonumber \\
&&(g^{\phi\phi}+\Gamma^2
  (v^\phi)^2)(\partial_\phi T)^2 +
2\Gamma^2 v^r v^\phi\partial_r T
  \partial_\phi T - \nonumber \\
&&\frac{2T}{e_{\rm eff} +P_{\rm eff}}\bigg((g^{rr}+\Gamma^2 (v^r)^2)
  \partial_r P_{\rm eff}\,\partial_r T+ \nonumber \\
&&(g^{\phi\phi}+\Gamma^2
  (v^\phi)^2)\partial_\phi T
  \,\partial_\phi P_{\rm eff} +
  \nonumber \\
&&\Gamma^2 v^r v^\phi\partial_r
  P_{\rm eff}
  \partial_\phi T +  \Gamma^2 v^r v^\phi\partial_\phi
  P_{\rm eff}
  \partial_r T\bigg)+\nonumber
\\
&&+\left(\frac{T}{e_{\rm eff} + P_{\rm eff}}\right)^2\bigg((g^{rr}+\Gamma^2 (v^r)^2)
  (\partial_r P_{\rm eff})^2+ \nonumber \\
&&(g^{\phi\phi}+\Gamma^2
  (v^\phi)^2)(\partial_\phi P_{\rm eff})^2 +
  \nonumber \\
&&2\Gamma^2 v^r v^\phi\partial_r
  P_{\rm eff}
  \partial_\phi P_{\rm eff} \bigg)
\bigg]\,.
\eea
The conversion of $\nabla_\mu \mathcal{S}^\mu$ from geometrized units to
 ${\rm cgs}$ units is given by
\be
\label{conv_rho}
[\nabla_\mu
\mathcal{S}^\mu]_{\rm{cgs}}= 1.0353 {\times} 10^{31}\,
G\,c\,\left(\frac{M_{\odot}}{M}\right)^3\,
[\nabla_\mu \mathcal{S}^\mu]_{\rm{geo}}\,.
\ee
In the code we generally compute the luminosity as the surface integral over
outgoing radiation fluxes $f^r_{\rm r}$ as
\be
L= 2\,\sum_{n=1}^{N_\phi} \left[\sqrt{\gamma}\,\left(f^r_{\rm r}\right)_n \,
\Delta \phi_n\right]\vert_{\tau=\tau_\bullet}\,,
\label{eq:luminosity}
\ee
where $\Delta \phi_n$ is the angular size of a grid cell and the integral
is taken at the radial position of the last
optically-thick surface\footnote{The factor $2$ in
  \eqref{eq:luminosity} accounts
for both the contributions above and below the equatorial
plane.}, i.e. where $\tau=\tau_\bullet$. 
In paper{\sc I} we computed the luminosity 
by imposing the criterion $\tau_\bullet\ge1$.
However, these small values of the optical depth often
correspond to an integration surface close to
the boundary of the numerical domain, where spurious boundary
effects may alter the results. Hence, in this paper we
have adopted a different criterion by choosing
$\tau_\bullet\ge10$, which guarantees that
the integration surface is not placed at the outermost
grid cells. 
For clarification we have repeated the luminosity
extraction for two models considered in paper{\sc I},
${\tt p.V18.cs07}$ and
${\tt p.V10.cs07}$, and show the light
curves, computed with the two different criteria, 
in Fig.~\ref{fig:Lcomp}.

\begin{figure}
{\includegraphics[clip=true,width=8.0cm]{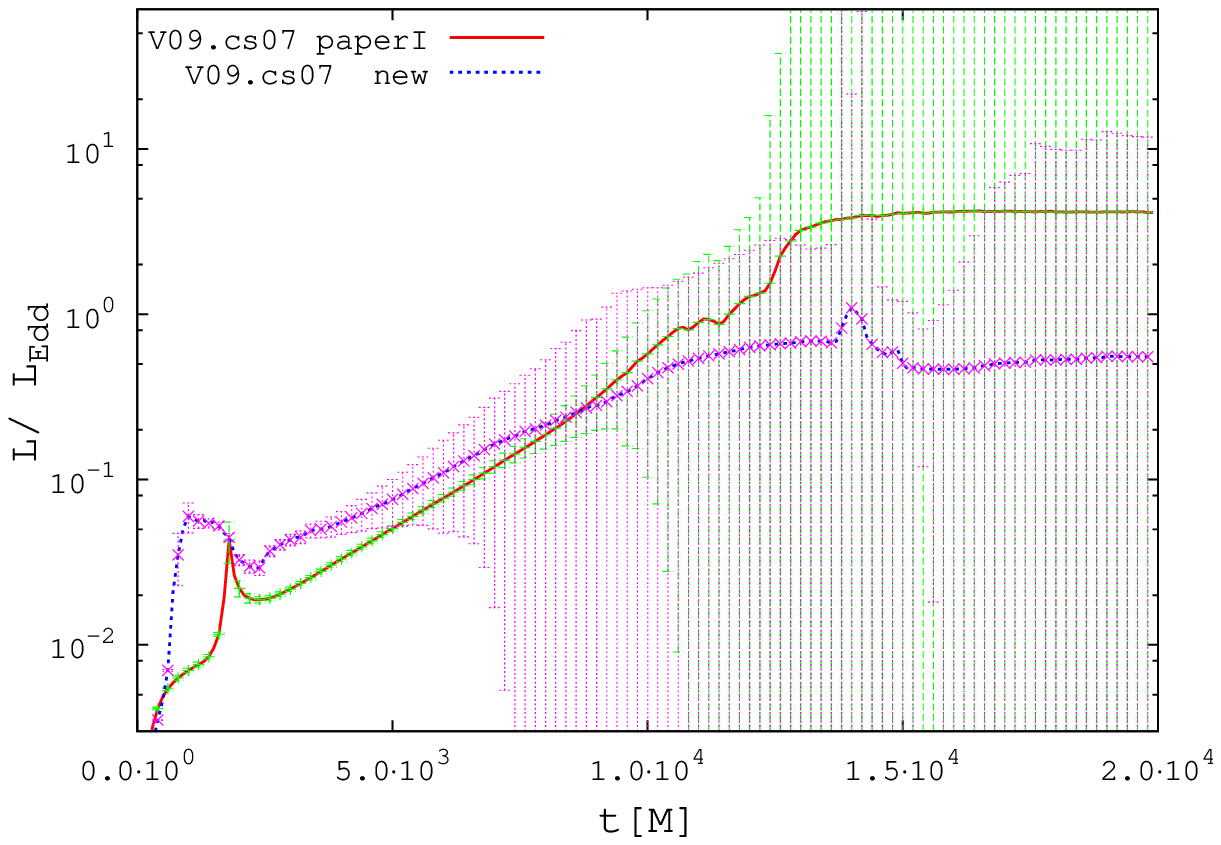}}
\caption{{\sc Uncertainties and comparison of luminosity extraction} - \textcolor{black}{${\tt V09.cs07}$}:
Extracting the luminosity at \textcolor{RubineRed}{$\tau\ge10$} leads to different light curves with much smaller uncertainties, these curves
are labelled "new". In paper{\sc I} we had used the criterion \textcolor{green}{ $\tau\ge1$}.
\label{fig:Lcomp2}
}
\end{figure}

We can assign error bars to our extraction method by
taking the standard deviation of the mean. 
For model ${\tt p.V09.cs07}$, the comparison
is shown, including the errorbars, in Fig.~\ref{fig:Lcomp2}. 
We stress that
the size of such error bars reflects the uncertainty in choosing the
position of the last optically thick surface
across which 
the emitted luminosity is computed.
It should be noted, moreover, that both our estimates agree within
these uncertainties, but  the choice $\tau\ge10$ produces 
much smaller error bars than $\tau\ge1$ and should
therefore be preferred. 


\section[]{Implementation of the IMEX scheme}
\label{appendixB}

A tableau notation is usually adopted to express in a
compact form the coefficients of the matrices $a_{ij}$,
$\tilde a_{ij}$ and
of the corresponding vectors $\omega_i$, $\tilde\omega_i$ as
\be
\begin{tabular} {c c c}
${c}$  & \vline & $a_{ij}$  \\
\hline 
              & \vline & ${\omega}^T$  \,,
\label{butcher_tableau}\end{tabular}
\ee
where the index $T$ denotes transposition
\footnote{Note
that the coefficients $c_i$ and $\tilde c_i$, which are
defined as $c_i=\sum_{j=1}^i a_{ij}$ and
 $\tilde c_i=\sum_{j=1}^i \tilde a_{ij}$ are not used in the
practical implementation of the scheme.}.

The explicit tableau of the SSP3$(4,3,3)$ is 
\be 
\begin{tabular} {c c c c c c}
 $0  $ & \vline & $0$  & $ 0 $ & $ 0 $ & $0$  \\
 $0  $ & \vline & $0$  & $ 0 $ & $ 0 $ & $0$  \\
 $1  $ & \vline & $0$  & $ 1 $ & $ 0 $ & $0$  \\
 $1/2$ & \vline & $0$  & $1/4$ & $1/4$ & $0$  \\
\hline 
   & \vline &  $0$ & $1/6$ & $1/6$ & $2/3$ \\
\end{tabular}
\ee
while the corresponding implicit tableau is 
\be 
\begin{tabular} {c c c c c c}
 $q_1$   & \vline & $q_1$  &  $0$  &  $0$  & $0$  \\
 $0$        & \vline & $-q_1$  &  $q_1$  &  $0$  & $0$  \\
 $1$        & \vline & $0$  &  $1-q_1$  &  $q_1$  & $0$ \\
 $1/2$      & \vline & $q_2$  & $q_3$ & $1/2-q_1-q_2-q_3$ & $q_1$ \\
\hline 
   & \vline &  $0$ & $1/6$ & $1/6$ & $2/3$ \\
\end{tabular}
\ee 
with 
\begin{eqnarray}
 &q_1 \equiv 0.24169426078821\,, &q_2 \equiv 0.06042356519705\,, \nonumber \\
 &q_3 \equiv 0.12915286960590\,  \,. \nonumber
\end{eqnarray} 

The coefficients of the radiation matrix $A(\bm Y)$ of Eq.~\eqref{explicit_dec}
are given by
\bea
A_{11}&=& -\alpha\Gamma(\chi^t + \chi^s 4 W (1 - \Gamma^2)) \nonumber\\
A_{12}&=&  \alpha\Gamma v^x (\chi^t + \chi^s W (1 - 4 \Gamma^2))\nonumber\\
A_{13}&=&  \alpha\Gamma v^y (\chi^t + \chi^s W (1 - 4 \Gamma^2))\nonumber\\
A_{14}&=&  \alpha\Gamma v^z (\chi^t + \chi^s W (1 - 4 \Gamma^2))\nonumber\\
A_{21}&=&  - \alpha\Gamma v_x \left[\chi^t(1-4W)+2 \chi^s(W-1) \right]\nonumber\\
A_{22}&=&   - \alpha\Gamma v^x   v_x \left[ \chi^t(2W-1)+\chi^s (2-W)\right]-  \nonumber\\
&&\alpha(\chi^t + \chi^s)/\Gamma \nonumber\\
A_{23}&=&  - \alpha\Gamma v^y v_x \left[ \chi^t(2W-1)+\chi^s (2-W)\right] \nonumber\\
A_{24}&=&  - \alpha\Gamma v^z v_x \left[ \chi^t(2W-1)+\chi^s (2-W)\right] \nonumber\\
A_{31}&=&  - \alpha\Gamma v_y \left[\chi^t(1-4W)+2 \chi^s(W-1) \right]\nonumber\\
A_{32}&=&  - \alpha\Gamma v^x v_y \left[ \chi^t(2W-1)+\chi^s (2-W)\right] \nonumber\\
A_{33}&=&   - \alpha\Gamma v^y v_y\left[
  \chi^t(2W-1)+\chi^s (2-W)\right] -  \nonumber\\
&&\alpha(\chi^t + \chi^s)/\Gamma \nonumber\\
A_{34}&=&    - \alpha\Gamma v^z v_y \left[ \chi^t(2W-1)+\chi^s (2-W)\right] \nonumber\\
A_{41}&=&  - \alpha\Gamma v_z \left[\chi^t(1-4W)+2 \chi^s(W-1) \right]\nonumber\\
A_{42}&=&  - \alpha\Gamma v^x v_z \left[ \chi^t(2W-1)+\chi^s (2-W)\right] \nonumber\\
A_{43}&=&   - \alpha\Gamma v^y v_z\left[
  \chi^t(2W-1)+\chi^s (2-W)\right]  \nonumber\\
A_{44}&=&    - \alpha\Gamma v^z v_z \left[
  \chi^t(2W-1)+\chi^s (2-W)\right]-\nonumber  \\
&&\alpha(\chi^t + \chi^s)/\Gamma \,, \nonumber
\eea
where, just for convenience, we have specified the
spatial coordinates to $(x,y,z)$.

\bibliographystyle{mn2e}
\bibliography{biblio/aeireferences}

\end{document}